\documentclass[12pt]{article}
\usepackage{epsfig}
\usepackage{rotating}
\usepackage{color}
\def\br(#1,#2){\left\langle#1#2\right\rangle}
\def\sq(#1,#2){\left[#1#2\right]}
\def\s(#1,#2){s_{#1 #2}}
\def\t(#1,#2,#3){s_{#1 #2 #3}}

\begin{document}

\begin{titlepage}

\hspace*{\fill}\parbox[t]{5.2cm}
{
FERMILAB-PUB-08-357-T \\
UCLA/08/TEP/27 \\
CP3-08-38 \\
FSU-HEP-080731 \\
ILL-(TH)-08-04 \\
\today} \vskip 2cm
\begin{center}
{\Large \bf Associated Production of a $W$ Boson and One $b$ Jet} \\
\medskip
\bigskip\bigskip\bigskip\bigskip
{\large  {\bf J.~Campbell},$^1$
         {\bf R.~K.~Ellis},$^2$
         {\bf F.~Febres Cordero},$^3$
         {\bf F.~Maltoni},$^4$ \\ \bigskip
         {\bf L.~Reina},$^5$
         {\bf D.~Wackeroth}$^6$,
     and {\bf S.~Willenbrock}$^7$} \\
\bigskip\bigskip\medskip
$^{1}$Department of Physics and Astronomy, University of Glasgow \\
Glasgow G12 8QQ, United Kingdom \\
\bigskip
$^{2}$Theoretical Physics Department, Fermi National Accelerator Laboratory \\
P.~O.~Box 500, Batavia, IL\ \ 60510 \\
\bigskip
$^{3}$Department of Physics and Astronomy, UCLA \\
Los Angeles, CA\ 90095-1547 \\
\bigskip
$^{4}$Institut de Physique Th\'{e}orique and \\
Centre for Particle Physics and Phenomenology (CP3) \\
Universit\'{e} Catholique de Louvain\\[1mm]
Chemin du Cyclotron 2 \\
B-1348 Louvain-la-Neuve, Belgium \\
\bigskip
$^{5}$Physics Department, Florida State University, Tallahassee, FL\
\ 32306-4350 \\
\bigskip
$^{6}$Department of Physics, SUNY at Buffalo, Buffalo, NY\
\ 14260-1500 \\
\bigskip
$^{7}$Department of Physics, University of Illinois at Urbana-Champaign \\
1110 West Green Street, Urbana, IL\ \ 61801 \\ \bigskip
\end{center}
\vfill
\newpage

\begin{abstract}
We calculate the production of a $W$ boson and a single $b$ jet to
next-to-leading order in QCD at the Fermilab Tevatron and the CERN
Large Hadron Collider.  Both exclusive and inclusive cross sections
are presented.  We separately consider the cross section for jets
containing a single $b$ quark and jets containing a $b\bar b$ pair.
There are a wide variety of processes that contribute, and it is
necessary to include them all in order to have a complete
description at both colliders.
\end{abstract}

\end{titlepage}

\section{Introduction}\label{sec:intro}

Many signals of new physics at hadron colliders involve a weak
vector boson ($W,Z$) plus jets containing heavy quarks ($c,b$).  For
example, the top quark was discovered via the signal $W+4j$, with at
least one $b$ jet \cite{Abe:1995hr,Abachi:1995iq}. More recently,
evidence for single top production has been presented via the signal
$W+2j$, with at least one $b$ jet
\cite{Abazov:2006gd,Abazov:2008kt,Acosta:2004bs,CDF:single}.  The
Higgs boson could manifest itself via the same signal, from the
production process $q\bar q\to Wh$, followed by $h\to b\bar b$
\cite{Aaltonen:2007wx,Aaltonen:2008zx,:2007hk}.  The Higgs boson
could also appear in the signal $Z+2j$ with at least one $b$ jet,
via $q\bar q\to Zh$ \cite{Abazov:2007pj,Aaltonen:2008mi}.

Most calculations of the background processes that give rise to
$W,Z+nj$ ($n=1,2$), where one or more jets contain heavy quarks
($Q=c,b$), have been completed at next-to-leading order (NLO)
\cite{Bern:1996ka,Bern:1997sc}. There exist NLO calculations of $ZQ$
\cite{Campbell:2003dd}, $ZQ\bar Q$
\cite{Campbell:2000bg,FebresCordero:2008ci}, $ZQj$
\cite{Campbell:2005zv}, $Wc$ \cite{Giele:1995kr,Campbell:2008},
$Wb\bar b$ \cite{Ellis:1998fv,FebresCordero:2006sj}, and $Wbj$
\cite{Campbell:2006cu}. An obvious omission from this list is $Wb$,
that is, $W+1j$ with at least one $b$ jet at NLO. It is the goal of
this paper to fill in that gap.  This NLO calculation can be used to
normalize the cross section from a leading-order event generator
such as ALPGEN \cite{Mangano:2002ea} or MadEvent
\cite{Maltoni:2002qb}.

It may seem surprising that the NLO calculation of $Wb$ has not
already been done, given all the other NLO calculations listed
above.  The reason for this, as we will discuss shortly, is that it
is essential to do this calculation with a finite $b$-quark mass. In
contrast, most of the above-mentioned calculations were done with a
vanishing heavy-quark mass, with the justification that the quark
mass is negligible at high transverse momentum ($p_T$).  The ability
to do NLO calculations with a finite heavy-quark mass for this class
of processes was only recently demonstrated, for $Wb\bar b$, in
Ref.~\cite{FebresCordero:2006sj} (and for $Zb\bar b$ in
Ref.~\cite{FebresCordero:2008ci}).  We will use this calculation,
together with the NLO calculation of $Wbj$ \cite{Campbell:2006cu},
to generate the NLO calculation of $Wb$, including the effect of the
$b$-quark mass.

We discuss the details of the calculation in Section \ref{sec:Wb}.
We then present results in Section \ref{sec:results} and conclusions
in Section \ref{sec:conclusions}.

\section{$Wb$ at NLO}\label{sec:Wb}

\begin{figure}[ht]
\begin{center}
\epsfxsize=5cm \epsfbox{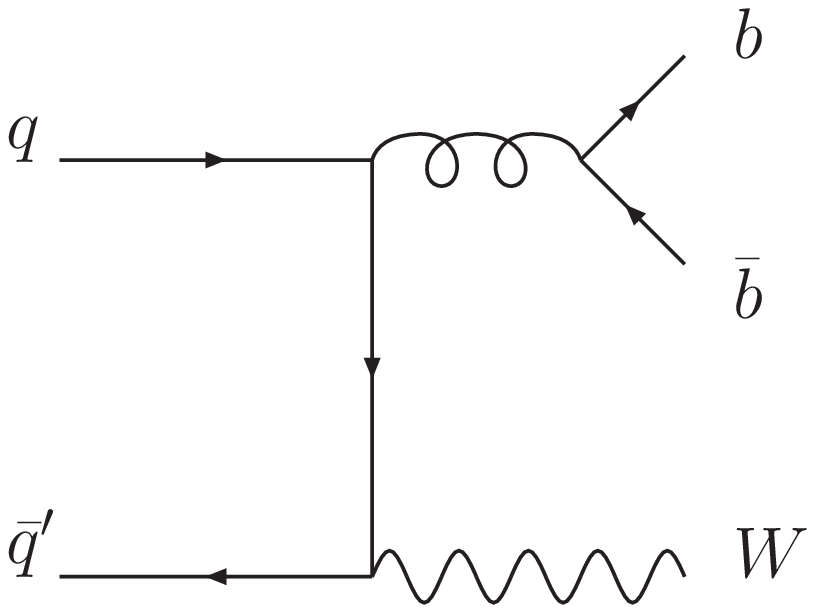} \hspace*{1cm}
\epsfxsize=5cm \epsfbox{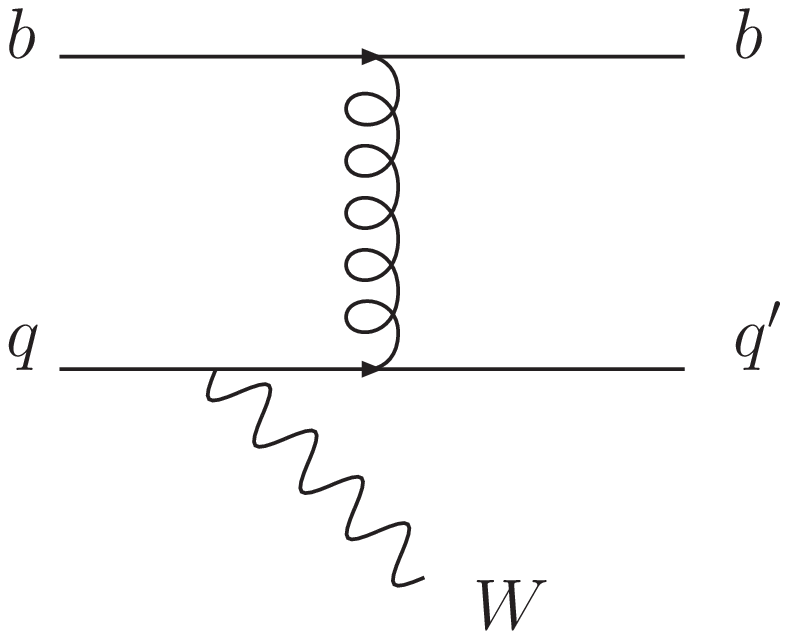} \\
(a) \hspace*{6cm} (b)
\end{center}
\caption{Leading-order processes for the production of a $W$ boson
and one jet, which contains a $b$ quark.} \label{fig:Wbb}
\end{figure}

The leading-order (LO) processes for the production of a $W$ boson
and one jet containing a $b$ quark are shown in Fig.~\ref{fig:Wbb}.
In both cases there are two partons in the final state, but we
require that only one of them (which contains a $b$ quark) reside at
high transverse momentum, with $p_{Tj}> 15$ GeV at the Fermilab
Tevatron ($\sqrt{s}=1.96$ TeV $p\bar p$) and $p_{Tj}>25$ GeV at the
CERN Large Hadron Collider (LHC, $\sqrt{s}=14$ TeV $pp$).  We also
require this parton to lie within a pseudorapidity of $|\eta_j|<2$
at the Tevatron and $|\eta_j|<2.5$ at the LHC.  Furthermore, we
demand that two partons be separated by $|\Delta R|>0.7$; if they
are not, then their four-momenta are added and they are considered
as occupying a single jet with this four-momentum.  These
requirements are made to crudely simulate the acceptance and
resolution of the detectors. They are listed in
Table~\ref{tab:parameters}, along with the parameters used in the
calculations.  In all calculations, the light quarks are summed over
$q=u,d,s,c$, including CKM mixing.

\begin{table} \begin{center} \caption[fake]{Cuts used
to simulate the acceptance and resolution of the detectors, and
parameters used in the calculations.}\label{tab:parameters}
\bigskip \begin{tabular} {ll}
Tevatron: $p_{Tj}>15$ GeV & $|\eta_j|<2$ \\
LHC: $\;\;\;\;\;\,$ $p_{Tj}>25$ GeV & $|\eta_j|<2.5$ \\
$|\Delta R_{b\bar b}|>0.7$ & $|\Delta R_{bj}|>0.7$ \\ \\
$M_W=80.44$ GeV & $m_b=4.7$ GeV \\
LO: CTEQ6L1 & NLO: CTEQ6M \cite{Pumplin:2002vw}\\
$\mu_F = M_W$ & $\mu_R = M_W$ \\
$\alpha_S^{LO}(M_Z)=0.130$ & $\alpha_S^{LO}(M_W)=0.132$ \\
$\alpha_S^{NLO}(M_Z)=0.118$ & $\alpha_S^{NLO}(M_W)=0.120$ \\
$g^2=8M_W^2G_F/\sqrt 2$ & $G_F=1.16639\times 10^{-5}$ GeV$^{-2}$ \\
$V_{ud}=V_{cs}=0.975$ & $V_{us}=V_{cd}=0.222$ \\
\end{tabular} \end{center}
\end{table}

\begin{figure}[ht]
\begin{center}
\vspace*{.2cm} \hspace*{0cm} \epsfxsize=7cm \epsfbox{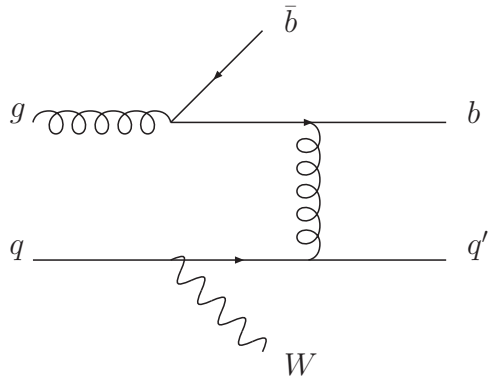}
\vspace*{-.8cm}
\end{center}
\caption{An alternative way of calculating the process in
Fig.~\ref{fig:Wbb}(b).} \label{fig:Wbbj}
\end{figure}

The LO process shown in Fig.~\ref{fig:Wbb}(b) contains a $b$ quark
in the initial state, and requires further discussion
\cite{Aivazis:1993pi,Collins:1998rz}.  To understand the usefulness
of this approach, consider the alternative approach, shown in
Fig.~\ref{fig:Wbbj} \cite{Mangano:2001xp}.  In that approach the
$b$-quark mass is kept nonzero, so one obtains a finite result even
when one of the $b$ quarks is emitted at zero $p_T$.  However,
although the collinear singularity is regulated by the $b$ mass, one
obtains an enhancement factor of $\ln M_W/m_b$. Another power of
this factor appears at each order of perturbation theory, degrading
the convergence of the series.  To ameliorate this, one sums this
enhancement factor to all orders into a $b$ distribution function,
and uses this function in the LO calculation of
Fig.~\ref{fig:Wbb}(b).  The other big advantage of this approach is
that the LO process of Fig.~\ref{fig:Wbb}(b) is simpler than that of
Fig.~\ref{fig:Wbbj}, and hence a NLO calculation becomes tractable.
This effectively allows to include a set of higher order
  corrections to the process of Fig.~\ref{fig:Wbbj} that would appear
only at NNLO in the fixed order calculation and will probably not be
available for quite some time.

There are a variety of processes that must be calculated at NLO:
\begin{enumerate}
\item $q\bar q'\to Wb\bar b$ at tree level [Fig.~\ref{fig:Wbb}(a)] and one
loop ($m_b\neq 0$)
\item $q\bar q'\to Wb\bar bg$ at tree level ($m_b\neq 0$)
\item $bq\to Wbq'$ at tree level [Fig.~\ref{fig:Wbb}(b)] and one
loop ($m_b=0$)
\item $bq\to Wbq'g$ at tree level ($m_b=0$)
\item $bg\to Wbq'\bar q$ at tree level ($m_b=0$)
\item $gq\to Wb\bar bq'$ at tree level (Fig.~\ref{fig:Wbbj})
($m_b\neq 0$)
\end{enumerate}
Processes 1 and 2 are calculated with a non-zero $b$-quark mass, using
the code developed in Ref.~\cite{FebresCordero:2006sj}.  Processes
3--5, which involve an initial-state $b$ quark, are calculated with
$m_b=0$, using the code developed in Ref.~\cite{Campbell:2006cu}.
Process 6, calculated with a non-zero $b$-quark mass, can be taken
from either code.  We used this process to cross-check the two
codes. We notice that the counterterm that subtracts from
  Process 6 the logarithmic terms already included in Process
  3 has been added, in our calculation, to Processes 3-5, since it
  shares the same final state phase space configuration.

\begin{figure}[ht]
\begin{center}
\vspace*{0cm} \hspace*{0cm}
\includegraphics[angle=90,width=14cm]{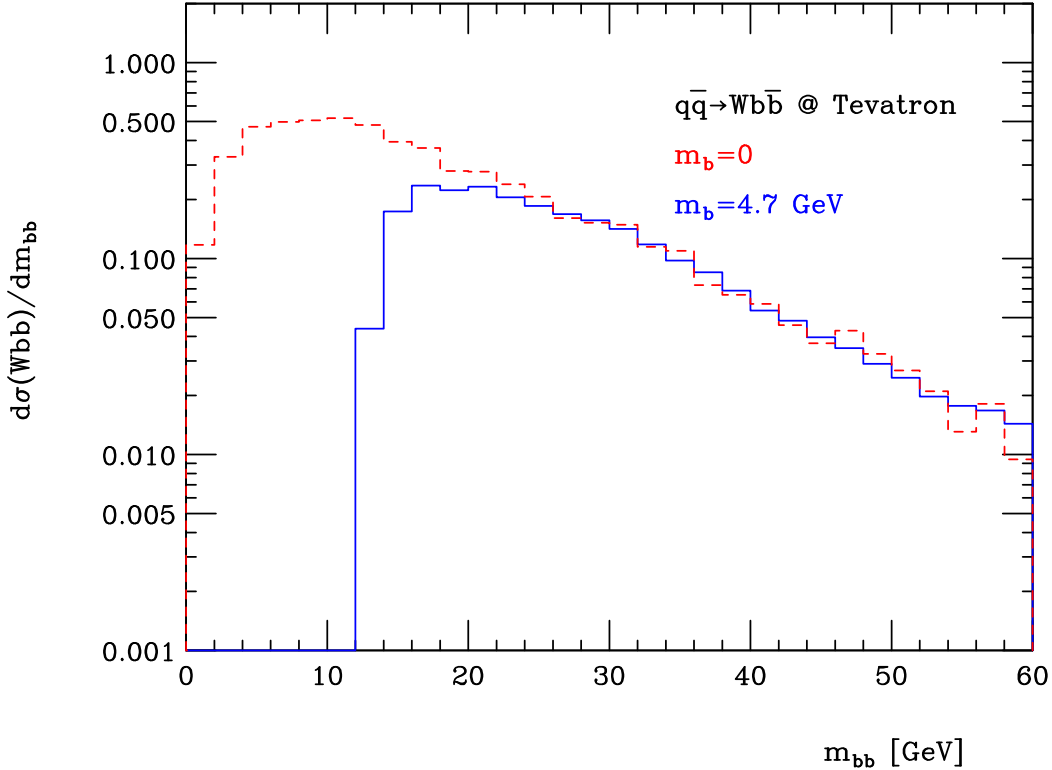}
\caption{The differential cross section for $W$ plus one $b$ jet
  ($p_{Tj}>15$ GeV, $|\eta_j|<2$), {\it vs}.\ the invariant mass of
  the $b$ quark and the other $b$ quark (outside the fiducial region and 
  separated by $\Delta R_{jj}>0.7$),
  at the Fermilab Tevatron ($\sqrt S=1.96$ TeV $p\bar p$).
  Only the contribution from the LO subprocess $q\bar q\to Wb\bar b$
  is shown. The (solid, blue) curve includes the $b$ quark mass, while
  the (dashed, red) curve does not.}\label{fig:tev-b}
\end{center}
\end{figure}

In the formalism that our calculation is based upon
\cite{Aivazis:1993pi,Collins:1998rz}, the $b$-quark mass is set to
zero in any process in which the $b$ quark appears in the initial
state (Processes 3--5).  This is not a limitation of the formalism,
nor is it an approximation.  In all other processes, where the $b$
quark appears only in the final state, the $b$-quark mass is kept
nonzero, although it is often a good approximation to set it to
zero. For the calculation we are performing, it is essential to keep
the $b$-quark mass nonzero.  This is because we demand that only one
$b$ jet be at high $p_T$, and we do not restrict the other $b$
quark. Thus there are regions of phase space where the $b\bar b$
invariant mass is not much greater than $2m_b$, in which case it is
very inaccurate to neglect the $b$-quark mass
\cite{Campbell:2006cu}. This issue arises already at leading order,
as evidenced by Fig.~\ref{fig:tev-b}, where we show the $b\bar b$
invariant mass from Process 1 for the $Wb$ exclusive cross section
at the Tevatron for both $m_b=0$ and $m_b=4.7$ GeV (with $\Delta
R_{jj}>0.7$). We see that the cross section (the area under the
curve) is very sensitive to the $b$-quark mass, even with the
required separation of the two $b$ quarks.

All NLO calculations are done in the $\overline {\rm MS}$ scheme.
When we refer to the NLO Processes 2, 4--6, it is understood that
initial-state collinear singularities are subtracted in this scheme.
For massless quarks, the collinear singularities are regulated
dimensionally, while for massive $b$ quarks (Processes 2 and 6) they are
regulated using a finite $b$ mass
\cite{Campbell:2006cu,Aivazis:1993pi,Collins:1998rz}.  The
calculation of Processes 3--6 was done using the Monte Carlo code
MCFM \cite{Ellis:1999ec,MCFM}.

\section{Results}\label{sec:results}

In Table~\ref{tab:WbXsecs_exc}, we give the exclusive cross section
for $W+1$ jet, where the jet contains a $b$ quark, and there are no
other jets present within the acceptance (listed in
Table~\ref{tab:parameters}).  We list the cross section for the case
where there is only one $b$ in the jet (denoted $Wb$), and when there
are two $b$'s in the jet (denoted $W(b\bar b)$).  The tagging
probability for a jet with two $b$'s differs from that of a jet with
one $b$ \cite{Acosta:2004nj}. We notice that in the
  $W(b\bar{b})$ case the two $b$ quarks can be collinear and give
  origin to large logarithms of the form $\ln(p_T^b/m_b)$.  Our
  results do not contain a resummation of these logarithms and are
  therefore subject to some degree of uncertainty.  We could have
  rejected the $W(b\bar{b})$ configuration in our $W+b$-jet NLO
  results or used a different jet algorith as suggested, for instance,
  in Ref.~\cite{Banfi:2007gu}. We prefer to keep it and quote it
  separately in order to make it available to different kinds of
  experimental analyses.  Table~\ref{tab:WbXsecs_exc} gives both the
LO cross section (in square brackets)\footnote{The
  contribution from Process 1 to the $Wb$ cross section at the
  Tevatron corresponds, at LO, to the area under the
  (solid, blue) curve in Fig.~\ref{fig:tev-b}.} and the NLO cross
section.  The first number given is from Processes 1--2; the second
from Processes 3--6; and the third is their sum.  In the
  case of $Wb$ we also quote, in parenthesis, the contribution of
  Process 6 by itself.  Indeed Process 6 is part of both the fixed
  order ($1+2+6$) and resummed ($3+4+5+6$)
  calculations\footnote{As explained in
    Section~\ref{sec:Wb}, Process 6 is the same in both the resummed
    and fixed order calculations, i.e. it does not include any
    counterterm to subtract the logarithmic terms that have been
    resummed in Process 3. Such counterterm has been included with
    Process 3 since it shares the same final state phase space.} and
  results could be quoted grouping Processes either way. In order to
  assess the impact of resumming initial state collinear logarithms,
  however, one needs to compare the sum of Processes 1,2, and 6 to the
  sum of process 3, 4, and 5. This can be easily deduced from our
  Tables knowing Process 6 independently. In the case of $W(b\bar
b)$, only Process 1 contributes at LO, and only Processes 1--2,6 at
NLO. Thus we do not need to quote Process 6 separately.
  The NLO results correspond to a pure fixed order calculation.  We
see that at the Tevatron, Processes 1--2 make a much larger
contribution to $Wb$ than Processes 3--6, while at the LHC the two
sets of processes make comparable contributions. This mirrors the
results for the $Wbj$ final state \cite{Campbell:2006cu}, first noted
in Ref.~\cite{Mangano:2001xp}. The fixed order
  calculation (Processes $1+2+6$) dominates at both the Tevatron and
  the LHC, but the corrections included in Processes 3+4+5 are as much
  as 25\% at the LHC and therefore very relevant. We also notice that
  for both $Wb$ and $W(b\bar b)$ Processes 1--2 dominate at the
Tevatron, while at the LHC Process 6 is nearly half the size of
Processes 1--2. Therefore it is very important to include
  higher order corrections to this process, as it is achieved by
  adding to the fixed order result Process 3, 4 and 5 (see discussion
  in Sec.~\ref{sec:Wb}).  Finally, we notice that for $W(b\bar b)$,
the NLO cross section is significantly larger than the LO cross
section, while the NLO correction is modest for $Wb$.

Also given in Table~\ref{tab:WbXsecs_exc} is the exclusive cross
section for $Wj$ ($j=u,d,s,c,g$), both at LO and NLO
\cite{Giele:1993dj,Campbell:2002tg}.  These numbers are useful to
compute the fraction of $W+1j$ events in which the jet contains a
$b$ quark. This fraction is around 0.7\% at the Tevatron and 0.8\%
at the LHC.

\begin{figure}[htp]
\begin{center}
\includegraphics*[scale=0.58,angle=-90]{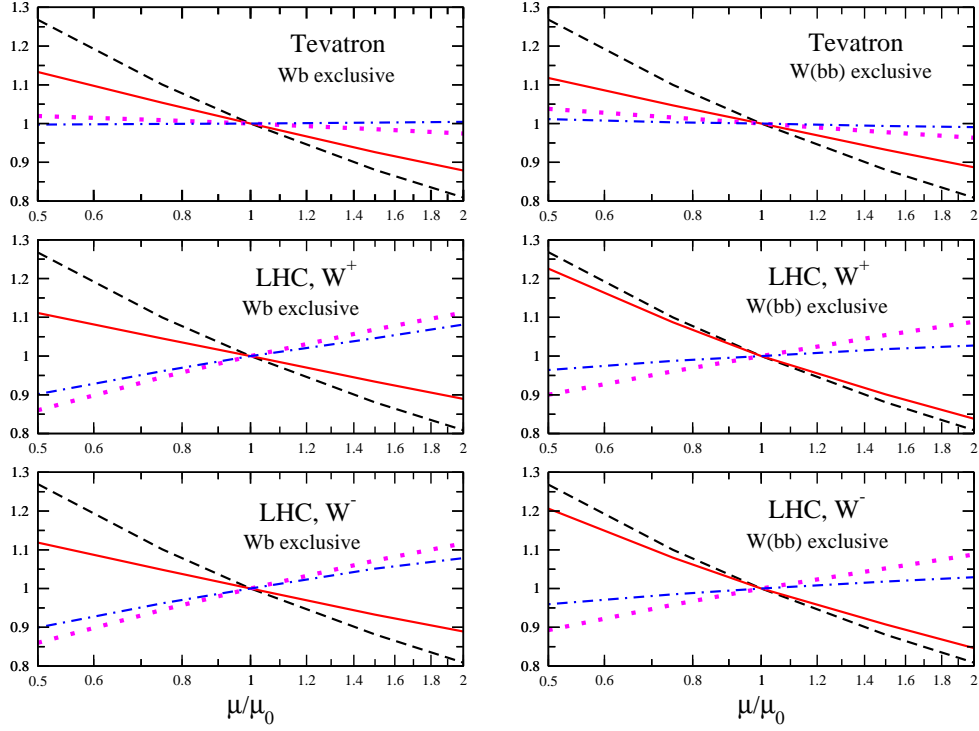}
\caption{Renormalization- and factorization-scale dependence of $Wb$
  (l.h.s.) and $W(b\bar{b})$ (r.h.s.) exclusive production. The upper
  plots correspond to both $W^+b/W^+(b\bar{b})$ and
  $W^-b/W^-(b\bar{b})$ production at the Tevatron. The middle plots
  correspond to $W^+b/W^+(b\bar{b})$ production at the LHC and the lower
  plots correspond to $W^-b/W^-(b\bar{b})$ production at the LHC. The
  black dashed (LO) and red solid (NLO) curves represent the
  dependence on the renormalization scale ($\mu_R$) when $\mu_R$ is
  varied with respect to its central value
  $\mu_0=M_W$, while the factorization scale ($\mu_F$) is fixed
  at $\mu_F=M_W$. In a similar way, the magenta dotted (LO) and blue
  dot-dashed (NLO) curves represent the dependence on $\mu_F$ when
  $\mu_F$ is varied with respect to its central
  value $\mu_0=M_W$, while $\mu_R$ is fixed at $\mu_R=M_W$. The
  cross sections have been normalized to their $\mu_R=\mu_F=M_W$
  value.}\label{fig:WbXsecs_exc_mudep}
\end{center}
\end{figure}

\begin{figure}[htp]
\begin{center}
\includegraphics*[scale=0.58,angle=-90]{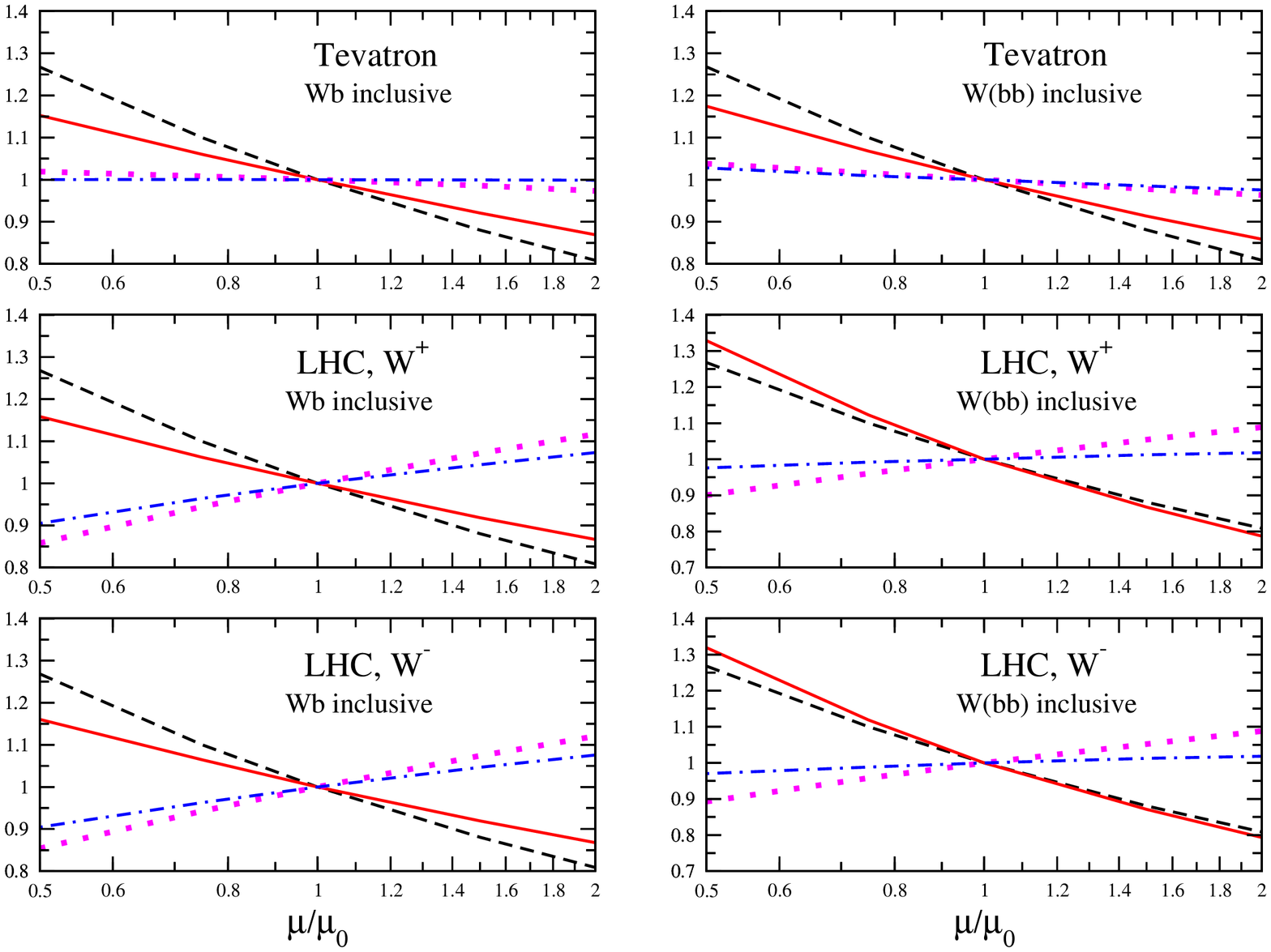}
\caption{Renormalization- and factorization-scale dependence of $Wb$
  (l.h.s.) and $W(b\bar{b})$ (r.h.s.) inclusive production. The upper
  plots correspond to both $W^+b/W^+(b\bar{b})$ and
  $W^-b/W^-(b\bar{b})$ production at the Tevatron. The middle plots
  correspond to $W^+b/W^+(b\bar{b})$ production at the LHC and the lower
  plots correspond to $W^-b/W^-(b\bar{b})$ production at the LHC. The
  black dashed (LO) and red solid (NLO) curves represent the
  dependence on the renormalization scale ($\mu_R$) when $\mu_R$ is
  varied with respect to its central value
  $\mu_0=M_W$, while the factorization scale ($\mu_F$) is fixed
  at $\mu_F=M_W$. In a similar way, the magenta dotted (LO) and blue
  dot-dashed (NLO) curves represent the dependence on $\mu_F$ when
  $\mu_F$ is varied with respect to its central
  value $\mu_0=M_W$, while $\mu_R$ is fixed at $\mu_R=M_W$. The
  cross sections have been normalized to their $\mu_R=\mu_F=M_W$
  value.}
\label{fig:WbXsecs_inc_mudep}
\end{center}
\end{figure}

We give in Table~\ref{tab:WbXsecs_inc} the inclusive cross sections
for $W+1j+X$, where the jet contains a $b$ quark, and where there may
be other jets present (up to two additional jets at NLO). The relative
importance of Processes 3--6 is significantly increased compared with
the exclusive cross sections, especially at the LHC, due
  in particular to Process 6. This is expected since Process 6 is more
  effectively cut in the exclusive cross section where the NLO cross
  section is required to have the same number of jets as the LO cross
  section. The fixed order cross section (Processes 1+2+6) dominates
  at both the Tevatron and LHC, but corrections coming from Processes
  3+4+5 are large and of the order of 50\% at the LHC. Moreover, given
  the relevance of Process 6, having included part of the NLO
  corrections to Process 6 by calculating Process 3 at NLO in QCD
  increases the stability and therefore the validity of the
  theoretical prediction. The NLO cross sections are also increased
by a larger factor than in the exclusive cross sections.  The fraction
of $W+1j+X$ events in which the jet contains a $b$ quark is around
0.9\% at the Tevatron and 1.2\% at the LHC.
\begin{table}[t]
\begin{center}
\caption{Exclusive cross sections (pb) for $W$ boson plus one jet,
  which contains at least one $b$ quark, at the Tevatron
  ($\sqrt{s}=1.96$ TeV $p\bar p$) and LHC ($\sqrt{s}=14$ TeV $pp$). No
  branching ratios or tagging efficiencies are included. The labels on
  the columns have the following meaning: $Wb=$ exactly one jet, which
  contains a $b$ quark; $W(b\bar b)=$ exactly one jet, which contains
  two $b$ quarks; $Wj=$ exactly one jet, including both light quarks
  ($u,d,s,c$) and gluons. Both the leading-order (in
  square brackets) and next-to-leading-order cross
  sections are given. The first number given is from Processes 1--2;
  the second number is from Processes 3--6 (with Process
    6 given separately in parenthesis); and the third number is their
  sum.}\label{tab:WbXsecs_exc}
\bigskip\begin{tabular}{|c|c|c|} \hline\hline
 & \multicolumn{2}{|c|}{Exclusive cross sections (pb)}\\
\hline
Collider & $Wb$ & $W(b\bar{b})$ \\
\hline
TeV $W^+(=W^-)$ & [5.28+0.75=6.03] 8.02+0.62(-0.05)=8.64 & [2.66] 3.73-0.02=3.71\\
LHC $W^+$ & [30.2+54.3=84.5] 40.0+48.4(22.6)=88.4 & [17.6] 22.7+11.7=34.4\\
LHC $W^-$ & [21.6+31.4=53.0] 29.8+29.4(12.6)=59.2 & [12.9] 17.2+6.5=23.7\\
\hline
 & \multicolumn{2}{|c|}{$Wj$}\\
\hline
TeV $W^+(=W^-)$ & \multicolumn{2}{|c|}{[1410] 1790}\\
LHC $W^+$ & \multicolumn{2}{|c|}{[14240] 15810}\\
LHC $W^-$ & \multicolumn{2}{|c|}{[11040] 12040}\\
\hline
\end{tabular}
\end{center}
\end{table}
\begin{table}[h]
\begin{center}
\caption{Inclusive cross sections (pb) for $W$ boson plus one jet,
  which contains at least one $b$ quark, at the Tevatron
  ($\sqrt{s}=1.96$ TeV $p\bar p$) and LHC ($\sqrt{s}=14$ TeV $pp$). No
  branching ratios or tagging efficiencies are included. The labels on
  the columns have the following meaning: $Wb+X=$ one or more jets, at
  least one of which contains a $b$ quark; $W(b\bar b)+X=$ one or more
  jets, one of which contains two $b$ quarks; $Wj+X=$ one or more
  jets, including both light quarks ($u,d,s,c$) and gluons. Both the
  leading-order (in square brackets) and
  next-to-leading-order cross sections are given. The first number
  given is from Processes 1--2; the second number is from Processes
  3--6 red (with Process 6 given separately in parenthesis);
  and the third number is their sum.}\label{tab:WbXsecs_inc}
\bigskip\begin{tabular}{|c|c|c|}
\hline\hline
 & \multicolumn{2}{|c|}{Inclusive cross sections (pb)}\\
\hline
Collider & $Wb+X$ & $W(b\bar{b})+X$ \\
\hline
TeV $W^+(=W^-)$ & [7.56+1.81=9.37] 11.77+2.40(0.77)=14.17 & [2.66] 4.17+0.39=4.56\\
LHC $W^+$ & [39.3+106.0=145.3] 53.6+136.1(68.9)=189.7  & [17.6] 25.1+35.9=61.0 \\
LHC $W^-$ & [27.9+67.0=94.9] 39.3+88.2(44.6)=127.5  & [12.9] 18.9+23.6=42.5 \\
\hline
 & \multicolumn{2}{|c|}{$Wj+X$}\\
\hline
TeV $W^+(=W^-)$ & \multicolumn{2}{|c|}{[1410] 2030}\\
LHC $W^+$ & \multicolumn{2}{|c|}{[14240] 20000}\\
LHC $W^-$ & \multicolumn{2}{|c|}{[11040] 15220}\\
\hline
\end{tabular}
\end{center}
\end{table}
%
\begin{table}[t]
\begin{center}
  \caption{Exclusive cross sections (pb) for $W$ boson plus one jet,
    which contains at least one $b$ quark, with the
    theoretical uncertainty due to renormalization (first uncertainty) and
    factorization (second uncertainty) scale dependence. The uncertainty due
    to the renormalization scale ($\mu_R$) dependence is estimated by
    varying $\mu_R$ by a factor of two with respect to its central
    value $\mu_R=M_W$, while keeping the factorization scale ($\mu_F$)
    fixed at its central value $\mu_F=M_W$. The uncertainty due to the
    factorization scale is estimated analogously, i.e. varying $\mu_F$
    by a factor of two about its central value $\mu_F=M_W$, while
    keeping $\mu_R$ fixed at $\mu_R=M_W$. The central values are
    extracted from Table~\ref{tab:WbXsecs_exc}. The labeling of
    columns and rows are as described in the caption of
    Table~\ref{tab:WbXsecs_exc}. Results within brackets are
    LO, results with no brackets are NLO.}\label{tab:WbXsecs_exc_mudep}
\begin{tabular}{|c|c|c|} \hline\hline
 & \multicolumn{2}{|c|}{Exclusive cross sections (pb)}\\
\hline
Collider & $Wb$ & $W(b\bar{b})$ \\
\hline TeV $W^+(=W^-)$ &
\big[6.03$\times(1^{+0.27+0.02}_{-0.19-0.03})$\big]
8.64$\times(1^{+0.13+0.004}_{-0.12-0.003})$ &
\big[2.66$\times(1^{+0.27+0.04}_{-0.19-0.04})$\big] 3.71$\times(1^{0.12+0.01}_{-0.11-0.01})$\\
LHC $W^+$ & \big[84.5$\times(1^{+0.27+0.11}_{-0.19-0.14})$\big]
88.4$\times(1^{+0.11+0.08}_{-0.11-0.10})$ &
\big[17.6$\times(1^{+0.27+0.09}_{-0.19-0.10})$\big] 34.4$\times(1^{+0.23+0.03}_{-0.16-0.04})$\\
LHC $W^-$ & \big[53.0$\times(1^{+0.27+0.12}_{-0.19-0.14})$\big]
59.2$\times(1^{+0.12+0.08}_{-0.11-0.10}$ &
\big[12.9$\times(1^{+0.27+0.09}_{-0.19-0.11})$\big] 23.7$\times(1^{+0.21+0.03}_{-0.15-0.04})$\\
\hline
\end{tabular}
\end{center}
\end{table}
\begin{table}[h]
\begin{center}
  \caption{Inclusive cross sections (pb) for $W$ boson plus one jet,
    which contains at least one $b$ quark, with the
    theoretical uncertainty due to renormalization (first uncertainty) and
    factorization (second uncertainty) scale dependence. The uncertainty due
    to the renormalization scale ($\mu_R$) dependence is estimated by
    varying $\mu_R$ by a factor of two with respect to its central
    value $\mu_R=M_W$, while keeping the factorization scale ($\mu_F$)
    fixed at its central value $\mu_F=M_W$. The uncertainty due to the
    factorization scale is estimated analogously, i.e. varying $\mu_F$
    by a factor of two about its central value $\mu_F=M_W$, while
    keeping $\mu_R$ fixed at $\mu_R=M_W$. The central values are
    extracted from Table~\ref{tab:WbXsecs_inc}. The labeling of
    columns and rows are as described in the caption of
    Table~\ref{tab:WbXsecs_inc}. Results within brackets are
    LO, results with no brackets are NLO.}\label{tab:WbXsecs_inc_mudep}
\begin{tabular}{|c|c|c|} \hline\hline
 & \multicolumn{2}{|c|}{Inclusive cross sections (pb)}\\
\hline
Collider & $Wb$ & $W(b\bar{b})$ \\
\hline
TeV $W^+(=W^-)$ & \big[9.37$\times(1^{+0.27+0.02}_{-0.19-0.03})$\big] 14.17$\times(1^{+0.15+0.0002}_{-0.13-0.001})$ & \big[2.66$\times(1^{+0.27+0.04}_{-0.19-0.04})$\big] 4.56$\times(1^{+0.17+0.03}_{-0.14-0.02})$\\
LHC $W^+$ & \big[145.3$\times(1^{+0.27+0.12}_{-0.19-0.14})$\big]
189.7$\times(1^{+0.16+0.07}_{-0.13-0.10})$  &
\big[17.6$\times(1^{+0.27+0.09}_{-0.19-0.10})$\big] 61.0$\times(1^{+0.33+0.02}_{-0.21-0.02})$ \\
LHC $W^-$ & \big[94.9$\times(1^{+0.27+0.12}_{-0.19-0.15})$\big]
127.5$\times(1^{+0.16+0.08}_{-0.13-0.10})$  &
\big[12.9$\times(1^{+0.27+0.09}_{-0.19-0.11})$\big] 42.5$\times(1^{+0.32+0.02}_{-0.21-0.03})$ \\
\hline
\end{tabular}
\end{center}
\end{table}
\newpage
We estimate the theoretical uncertainty in the exclusive and
inclusive LO and NLO cross sections by varying the renormalization
($\mu_R$) and factorization ($\mu_F$) scales independently. We vary
$\mu_R$ in the $(0.5-2)M_W$ range while keeping $\mu_F$ fixed at its
central value, $\mu_F=M_W$. Similarly we vary $\mu_F$ in the
$(0.5-2)M_W$ range keeping $\mu_R=M_W$. The results are illustrated
in Figs.~\ref{fig:WbXsecs_exc_mudep} and
\ref{fig:WbXsecs_inc_mudep}, where the plots on the l.h.s.\
correspond to $W$ production with a $b$ jet ($Wb$) while the plots
on the r.h.s.\ correspond to $W$ production with a double-$b$ jet
($W(b\bar{b})$).  The upper plots refer to
$W^+b/W^+(b\bar{b})=W^-b/W^-(b\bar{b})$ production at the Tevatron,
the middle plots to $W^+b/W^+(b\bar{b})$ production at the LHC, and
the lower plots to $W^-b/W^-(b\bar{b})$ production at the LHC.  The
cross sections in Figs.~\ref{fig:WbXsecs_exc_mudep} and
\ref{fig:WbXsecs_inc_mudep} have been normalized to their
$\mu_R=\mu_F=M_W$ value. The horizontal axis represents the
variation of either $\mu_R$ or $\mu_F$, depending on the curve (see
figure captions), normalized to the central value $\mu_0=M_W$.  In
Tables~\ref{tab:WbXsecs_exc_mudep} and \ref{tab:WbXsecs_inc_mudep}
we quantitatively give the variation with $\mu_R$ and $\mu_F$ as
(asymmetric) uncertainties on the central value, corresponding to
the choice $\mu_R=\mu_F=M_W$ used to obtain the results of
Tables~\ref{tab:WbXsecs_exc} and \ref{tab:WbXsecs_inc}.
We have not included in the theoretical uncertainties reported in
this paper the uncertainty coming from the parton distribution
functions.

From Figs.~\ref{fig:WbXsecs_exc_mudep} (and
\ref{fig:WbXsecs_inc_mudep} and Tables~\ref{tab:WbXsecs_exc_mudep}
and \ref{tab:WbXsecs_inc_mudep}) we see that the theoretical
uncertainty due to the dependence on the renormalization scale is
larger than the corresponding uncertainty from the
factorization-scale dependence.  The decrease in the
factorization-scale dependence in going from LO to NLO is mild,
while the decrease in the renormalization-scale dependence is
significant.  An exception is the inclusive $W(b\bar{b})$
cross-sections at the LHC, where the renormalization-scale
dependence slightly increases at NLO. Even in the exclusive case,
the improvement in the renormalization-scale dependence in going
from LO to NLO for $W(b\bar{b})$ is mild at the LHC.  These
exceptions can be explained by the fact that only Processes 1--2,6
contribute to $W(b\bar{b})$ production at NLO and, among those,
Process 6 opens a new initial state, namely $qg$, and is effectively
a LO process. The effect is larger at the LHC because, due to the
large gluon density, the $qg$ channel is more relevant. The effect
is also larger for inclusive rather than exclusive cross sections
because the exclusive final state suppresses the contribution of
Process 6 (which is a $2\rightarrow 4$ process), as evidenced by the
numerical results in Tables~\ref{tab:WbXsecs_exc} and
\ref{tab:WbXsecs_inc}.

\begin{figure}[htp]
\begin{center}
\begin{tabular}{cc}
\includegraphics*[scale=0.4,angle=90]{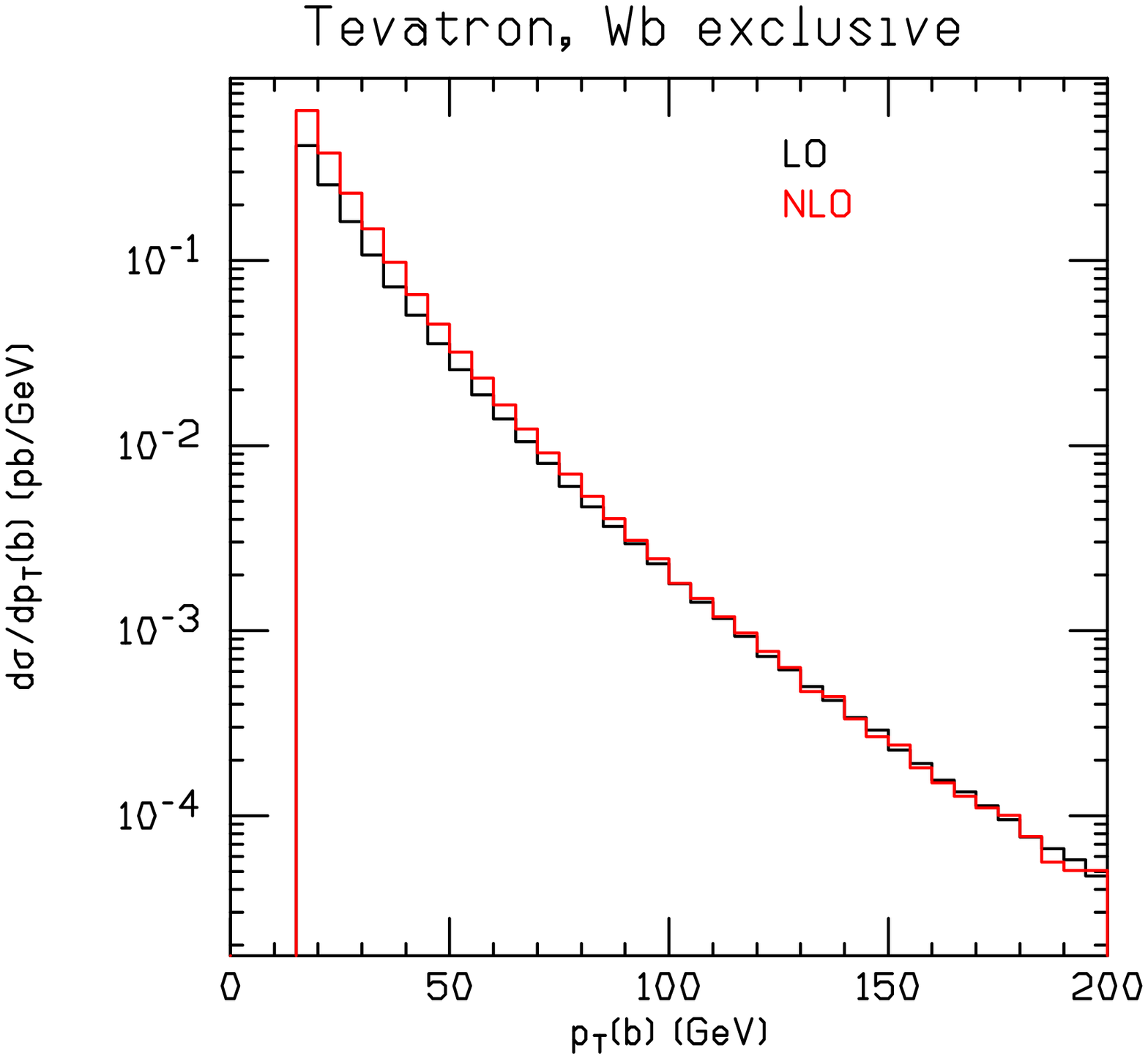} &
\includegraphics*[scale=0.4,angle=90]{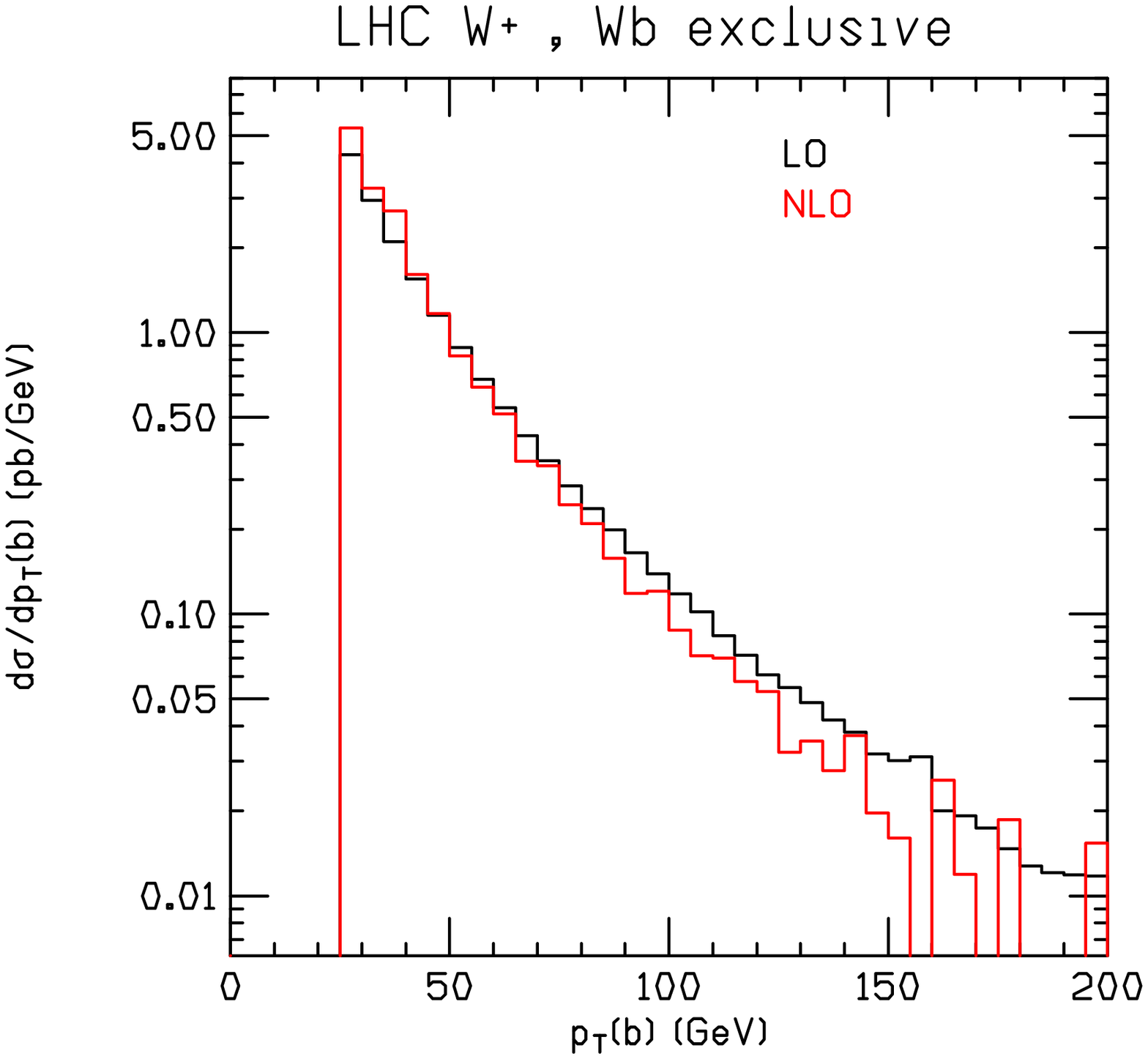}\\
\includegraphics*[scale=0.4,angle=90]{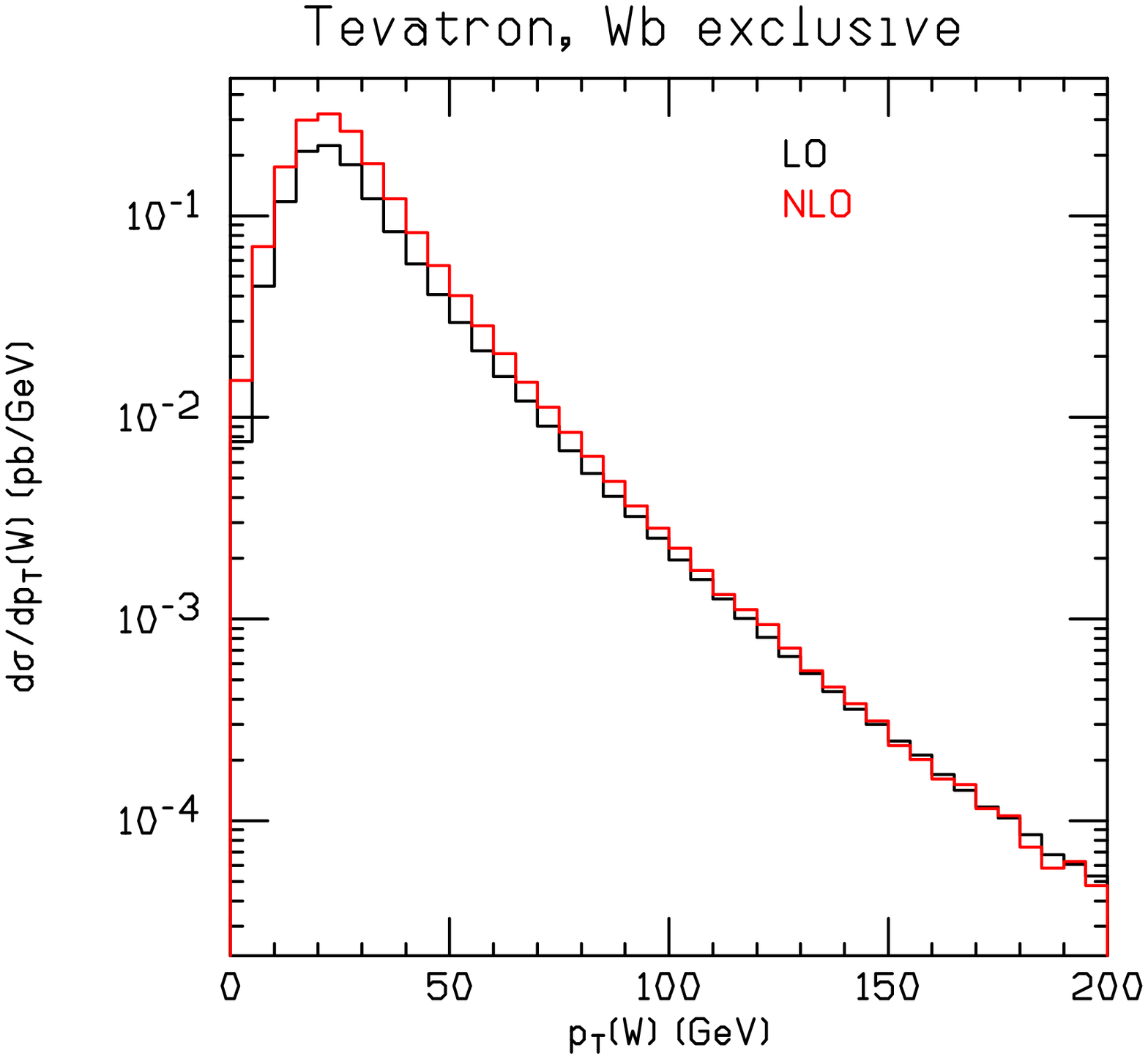} &
\includegraphics*[scale=0.4,angle=90]{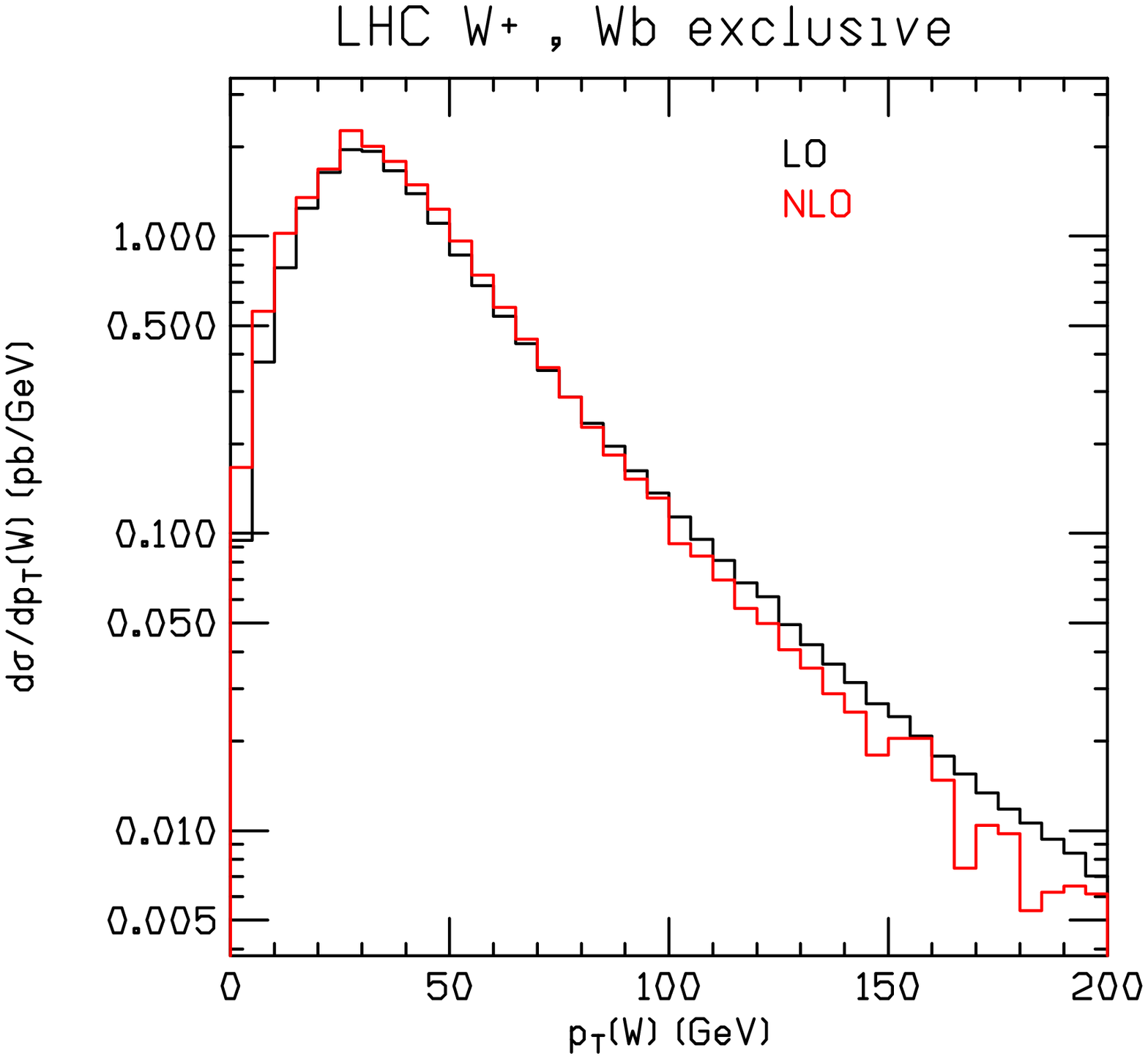}
\end{tabular}
\caption{$p_T(b)$ (upper plots) and $p_T(W)$ (lower plots)
  distributions for $W^+b$ \emph{exclusive} production, at the Tevatron
  (l.h.s.\ plots) and at the LHC (r.h.s.\ plots).  }
\label{fig:dist_signature3}
\end{center}
\end{figure}

\begin{figure}[htp] 
\begin{center} 
\begin{tabular}{cc}
\includegraphics*[scale=0.4,angle=90]{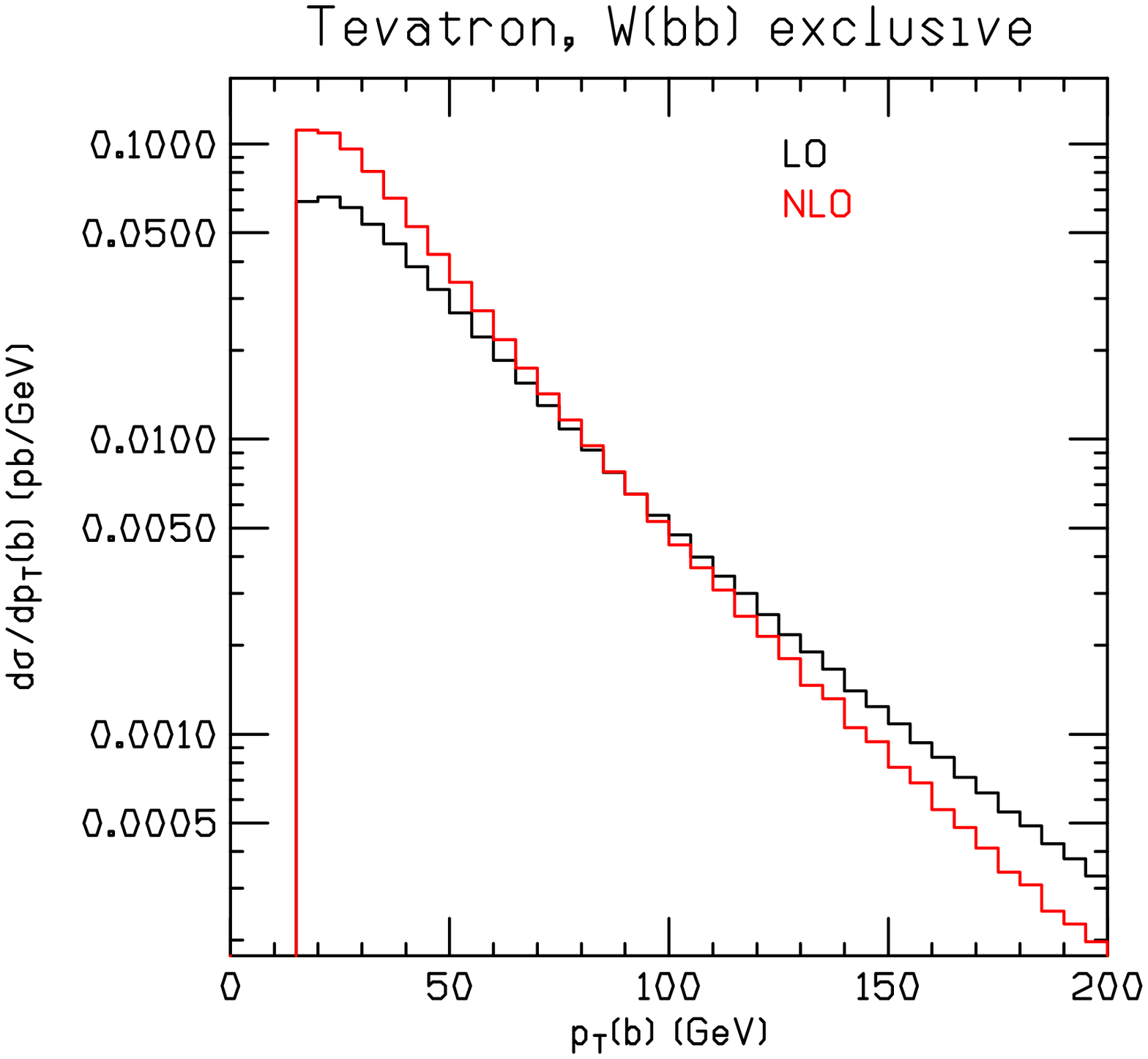}&
\includegraphics*[scale=0.4,angle=90]{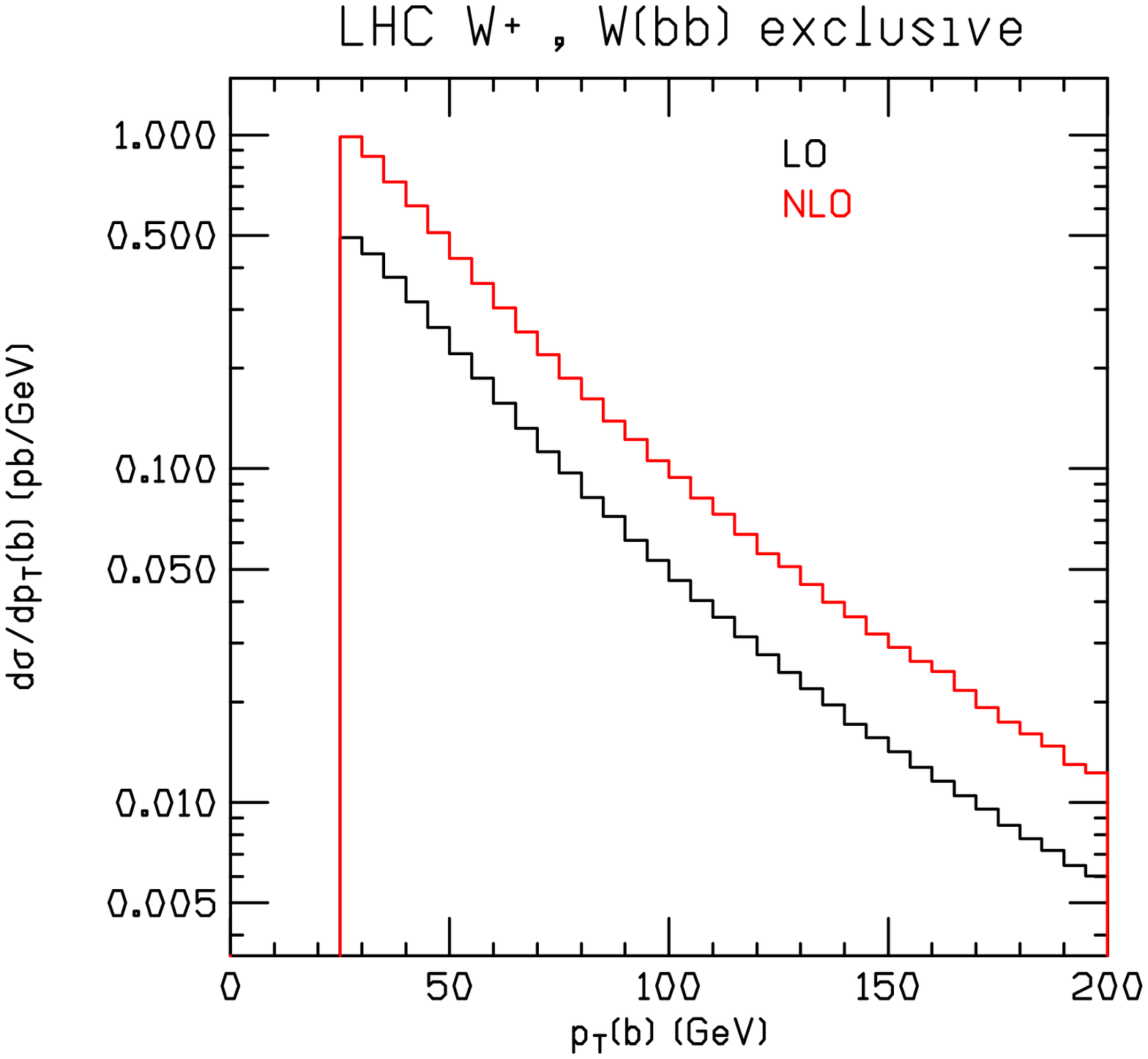}\\
\includegraphics*[scale=0.4,angle=90]{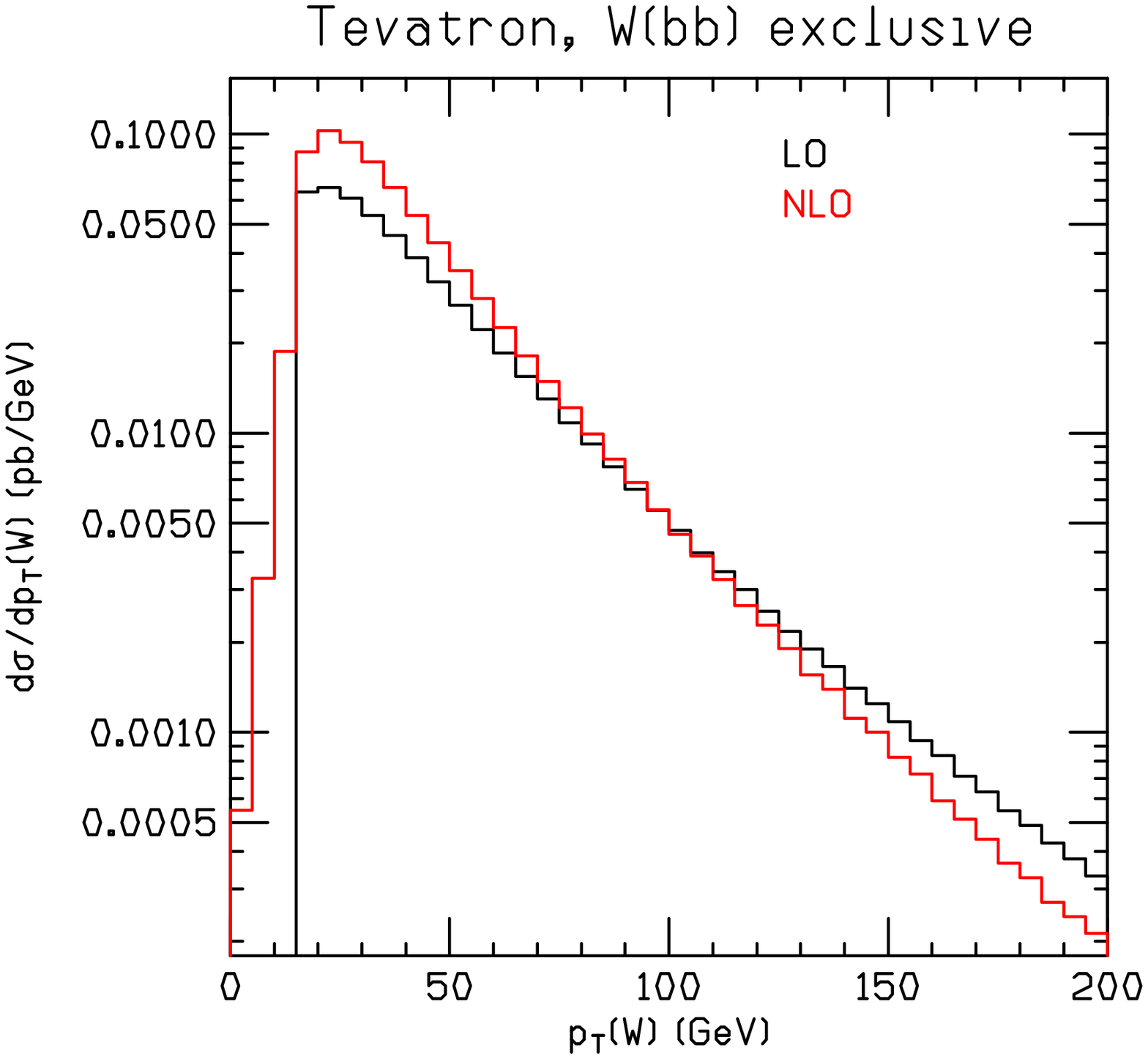}&
\includegraphics*[scale=0.4,angle=90]{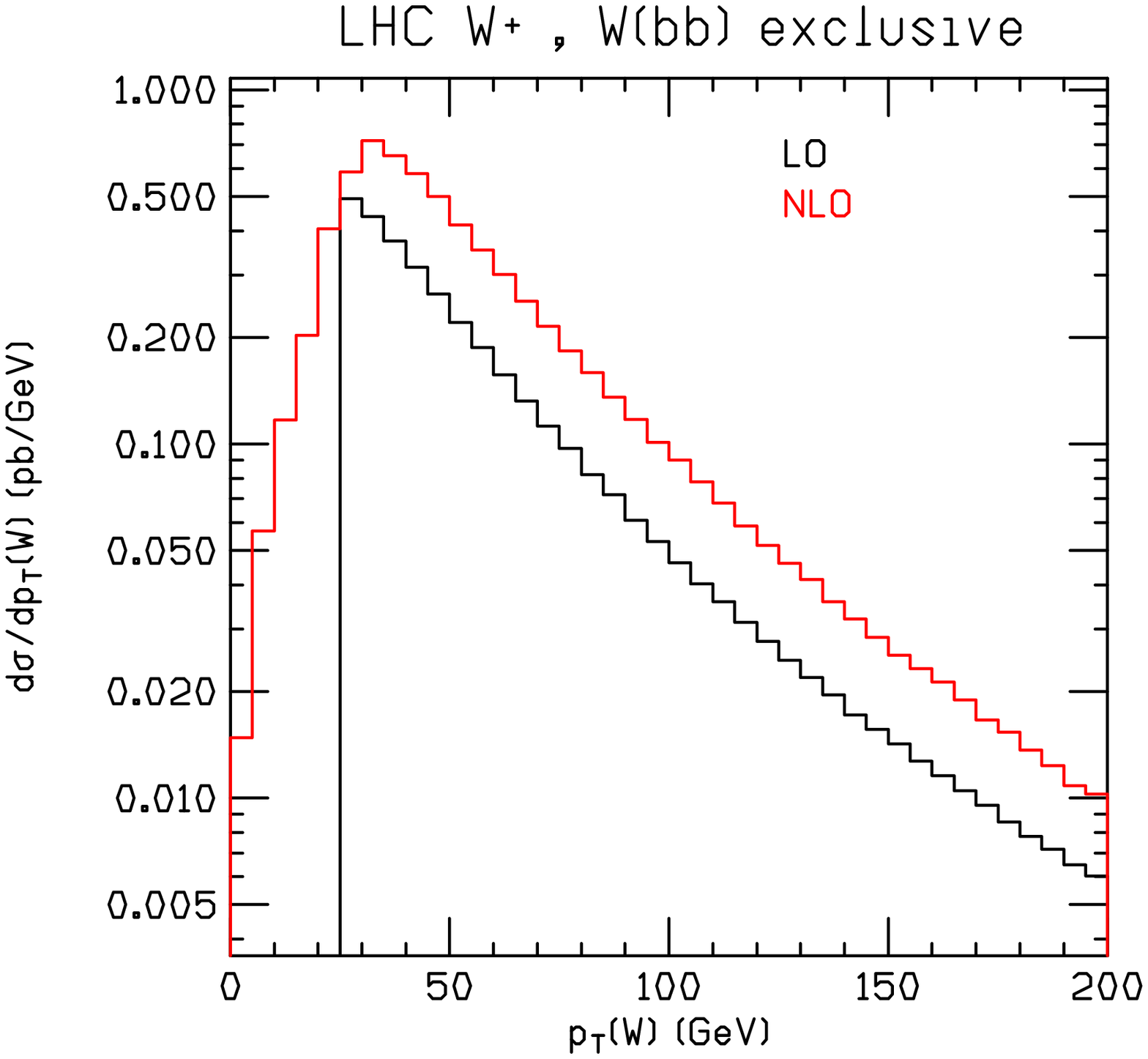}
\end{tabular}
\caption{$p_T(b)$ (upper plots) and $p_T(W)$ (lower plots)
distributions for $W^+(b\bar{b})$ \emph{exclusive} production, at
the Tevatron (l.h.s.\ plots) and at the LHC (r.h.s.\ plots).}
\label{fig:dist_signature4}
\end{center}
\end{figure}

\begin{figure}[htp]
\begin{center}
\begin{tabular}{cc}
\includegraphics*[scale=0.4,angle=90]{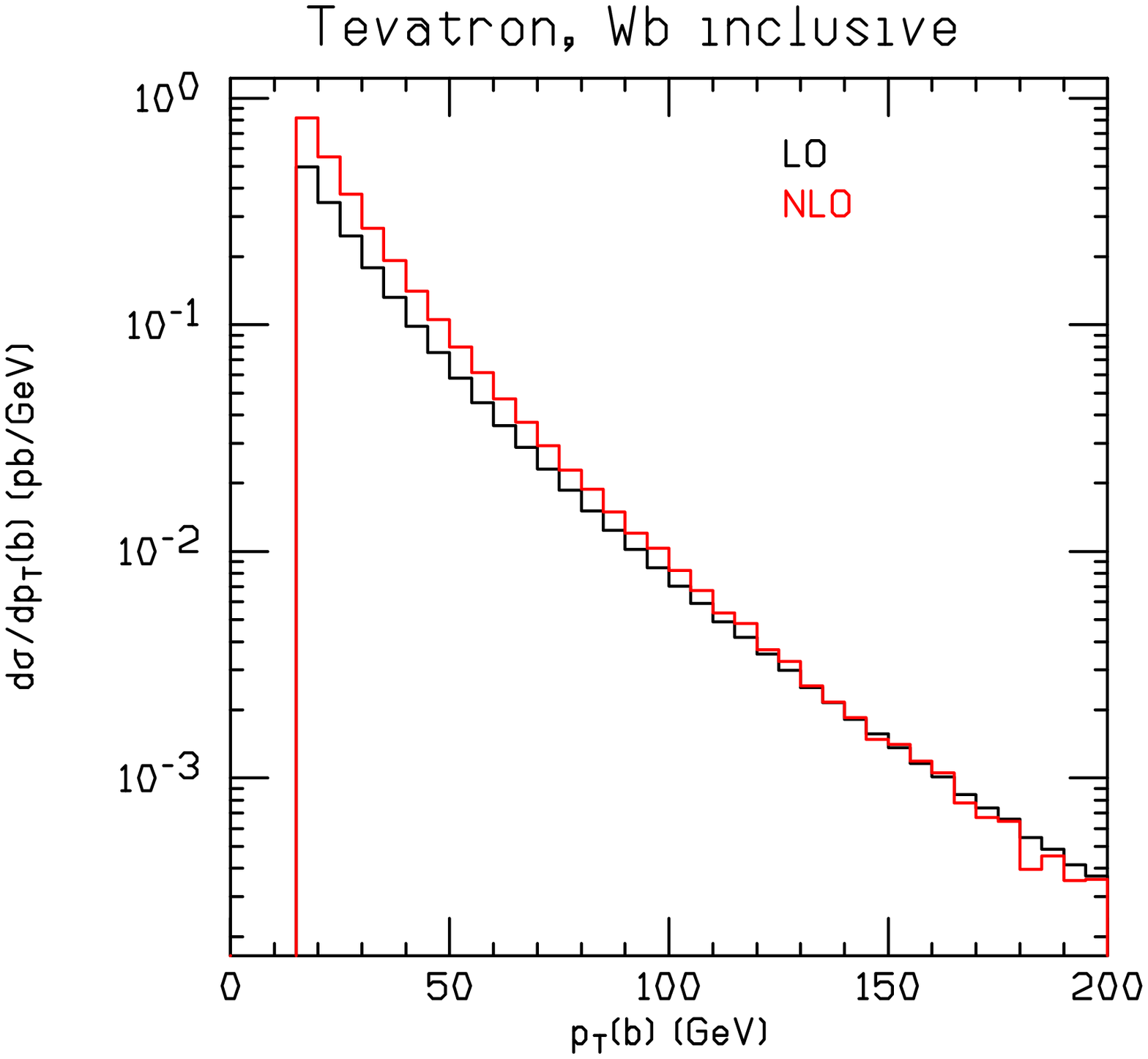} &
\includegraphics*[scale=0.4,angle=90]{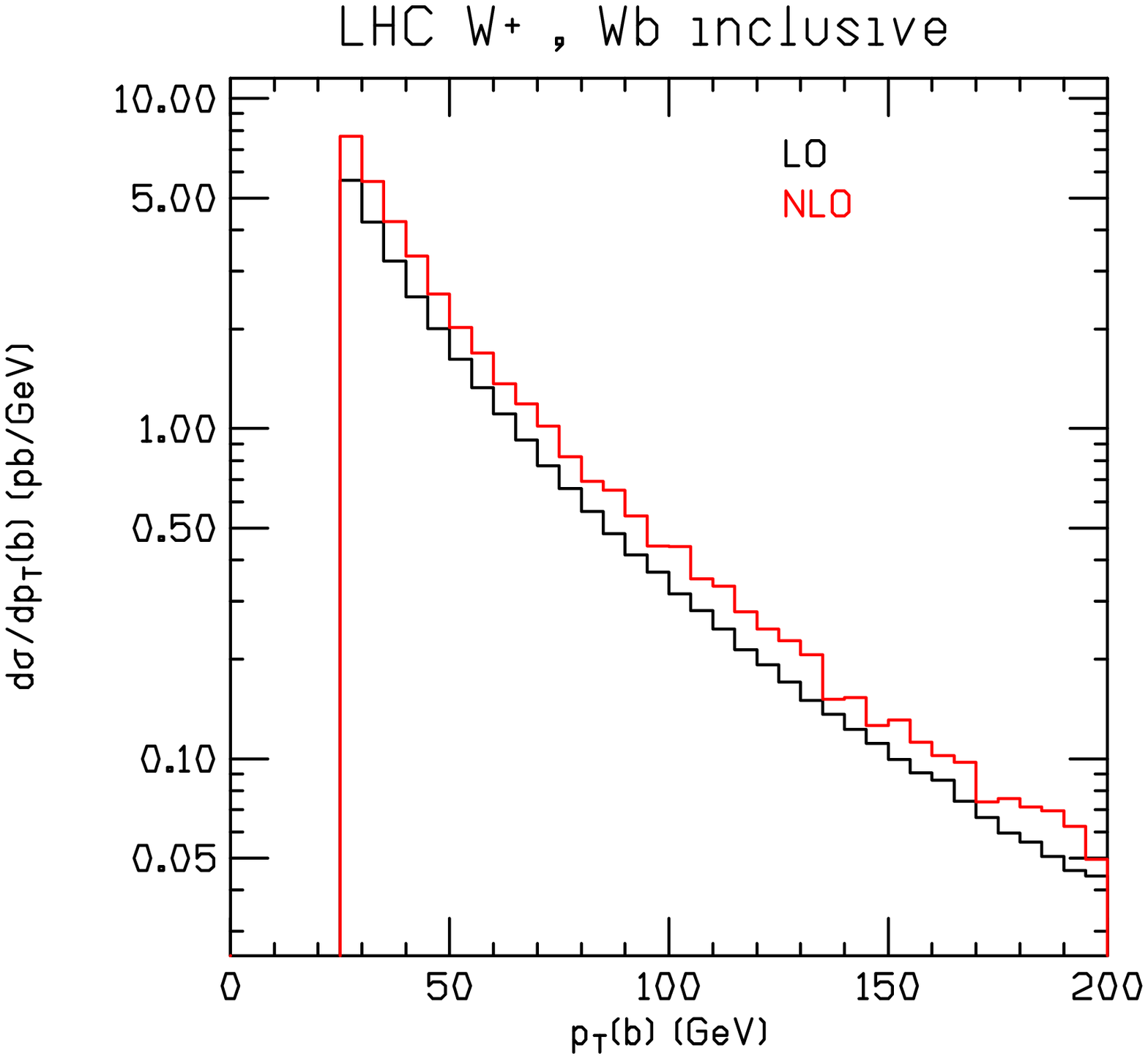}\\
\includegraphics*[scale=0.4,angle=90]{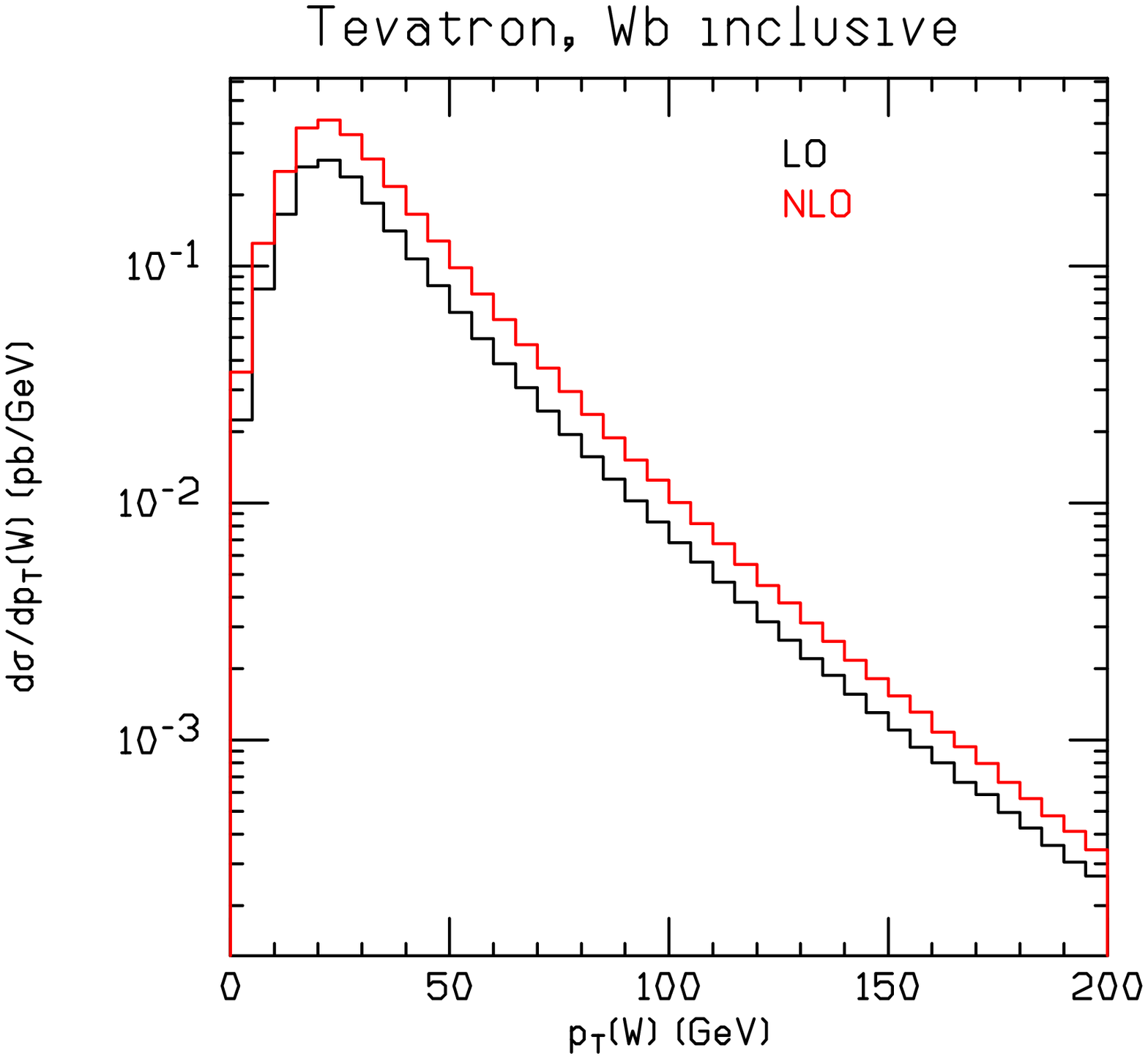} &
\includegraphics*[scale=0.4,angle=90]{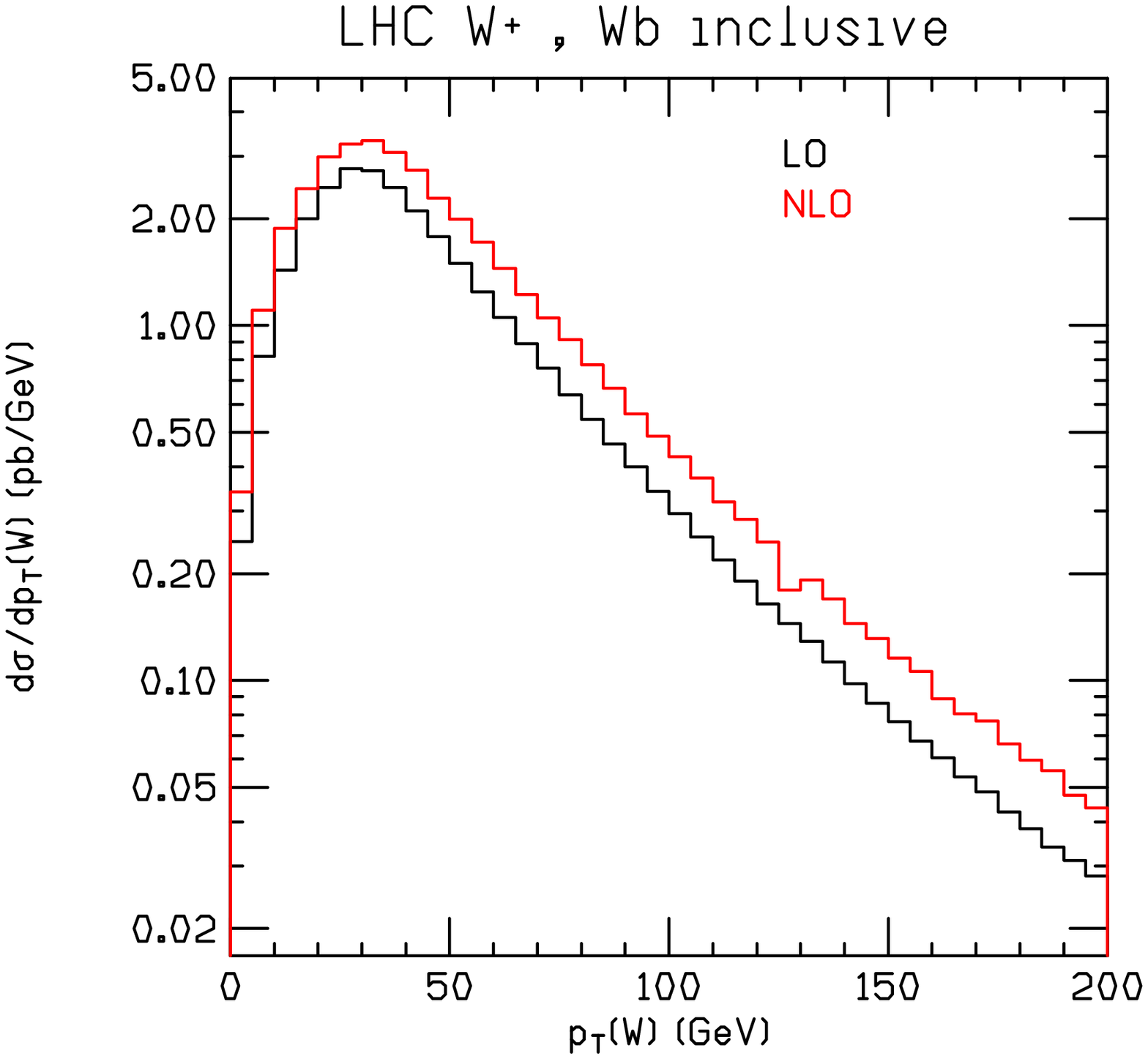}
\end{tabular}
\caption{$p_T(b)$ (upper plots) and $p_T(W)$ (lower plots)
  distributions for $W^+b$ \emph{inclusive} production, at the Tevatron
  (l.h.s.\ plots) and at the LHC (r.h.s.\ plots).  }
\label{fig:dist_signature1}
\end{center}
\end{figure}

\begin{figure}[htp]
\begin{center}
\begin{tabular}{cc}
\includegraphics*[scale=0.4,angle=90]{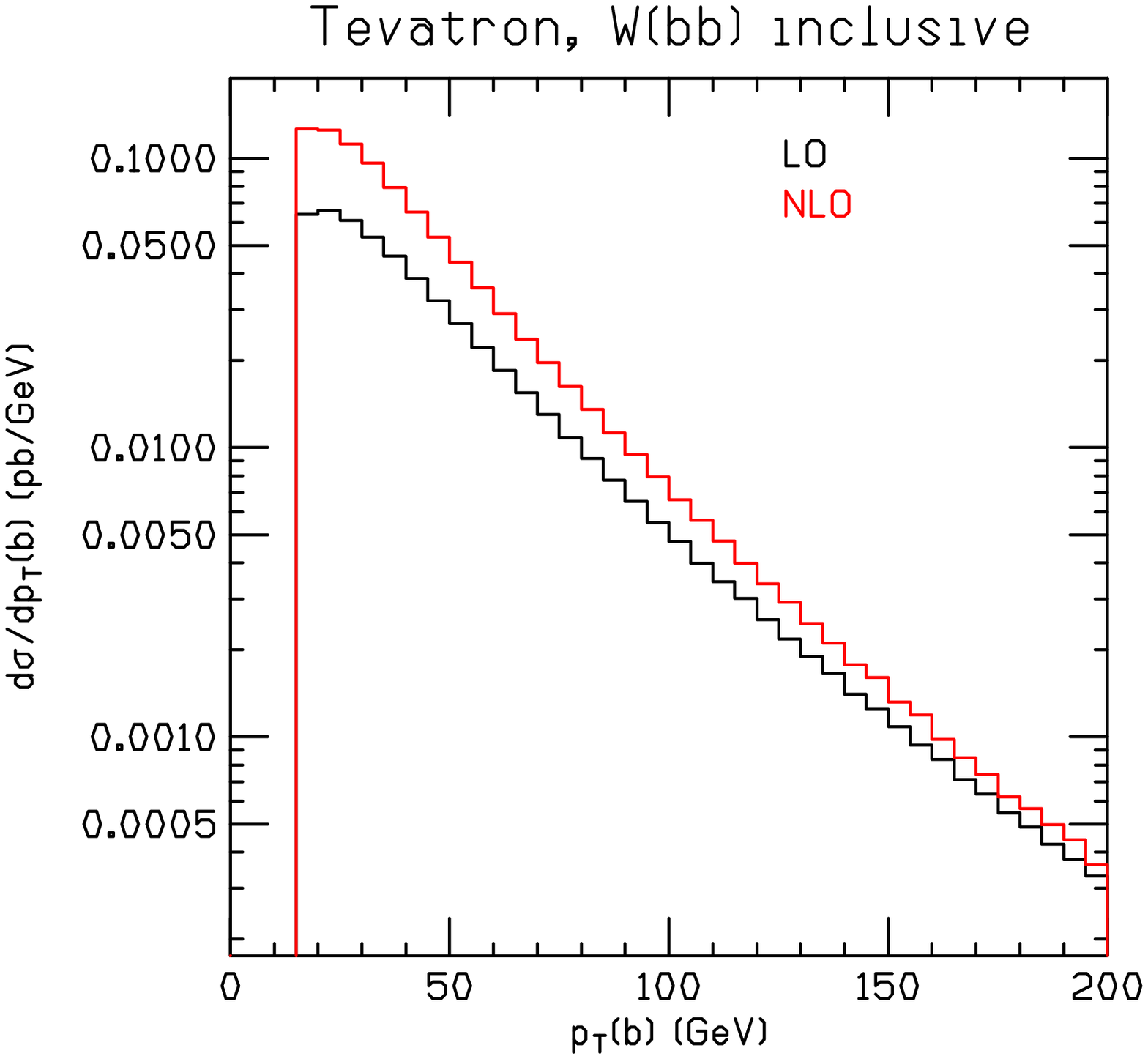} &
\includegraphics*[scale=0.4,angle=90]{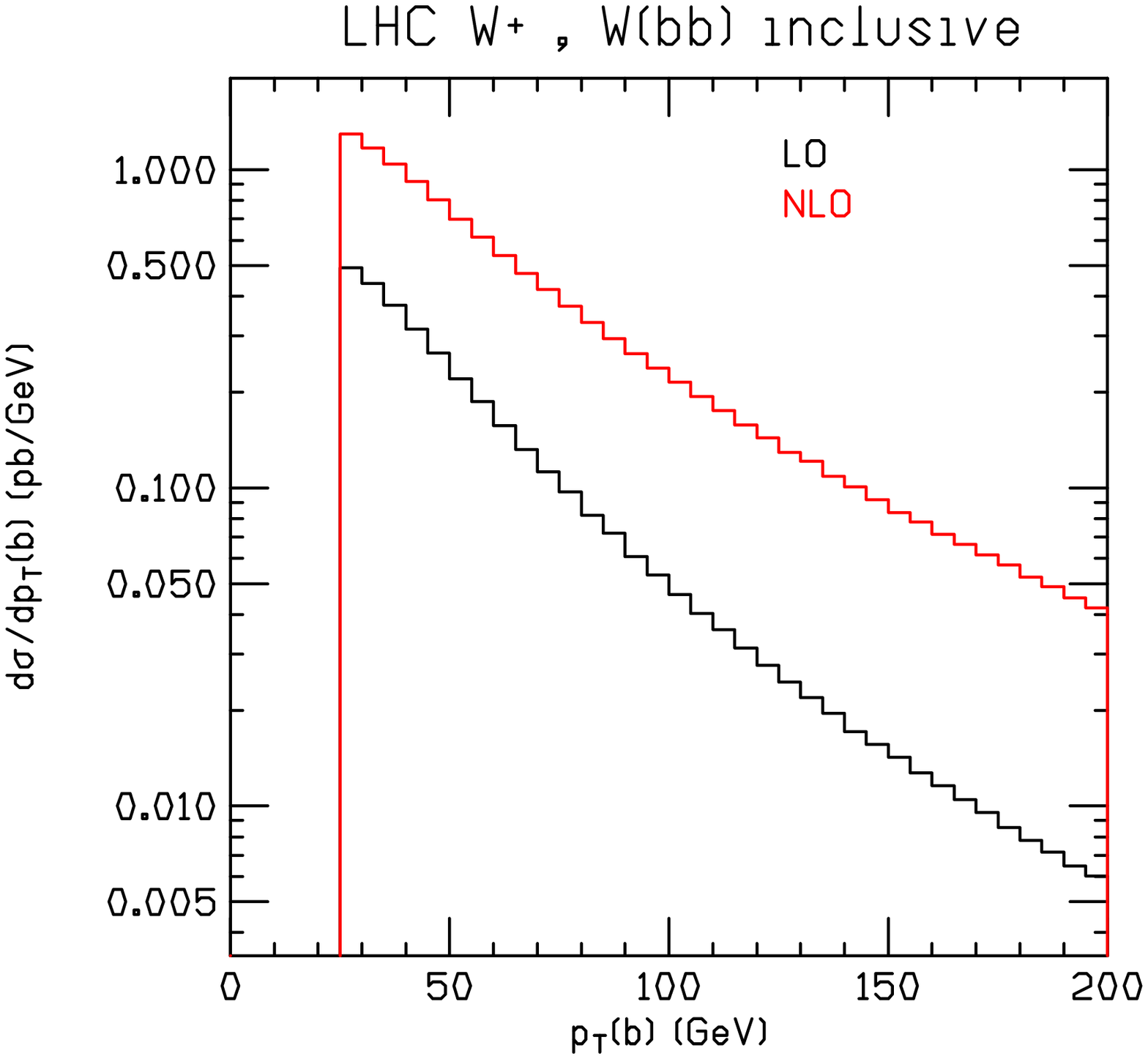}\\
\includegraphics*[scale=0.4,angle=90]{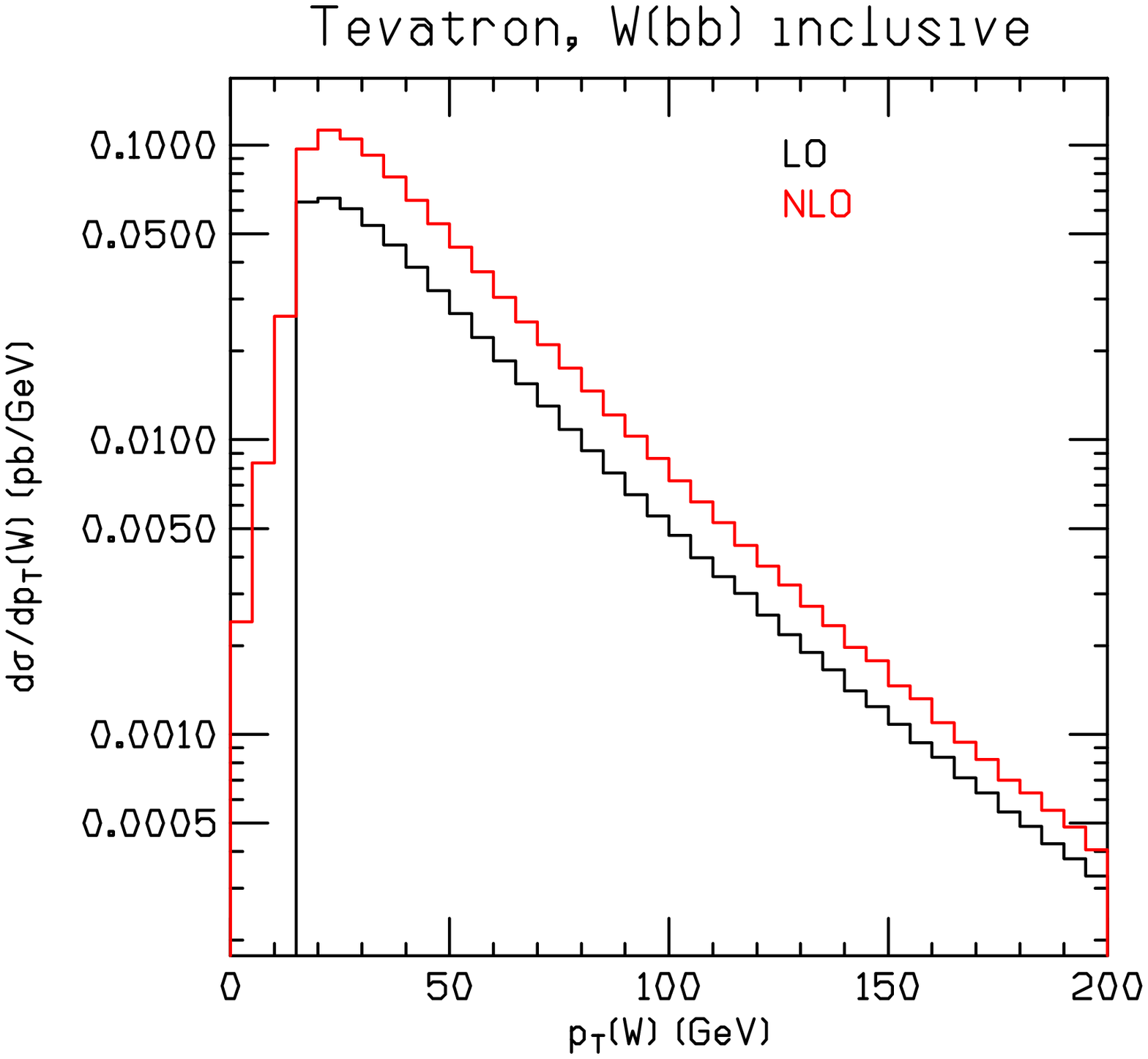} &
\includegraphics*[scale=0.4,angle=90]{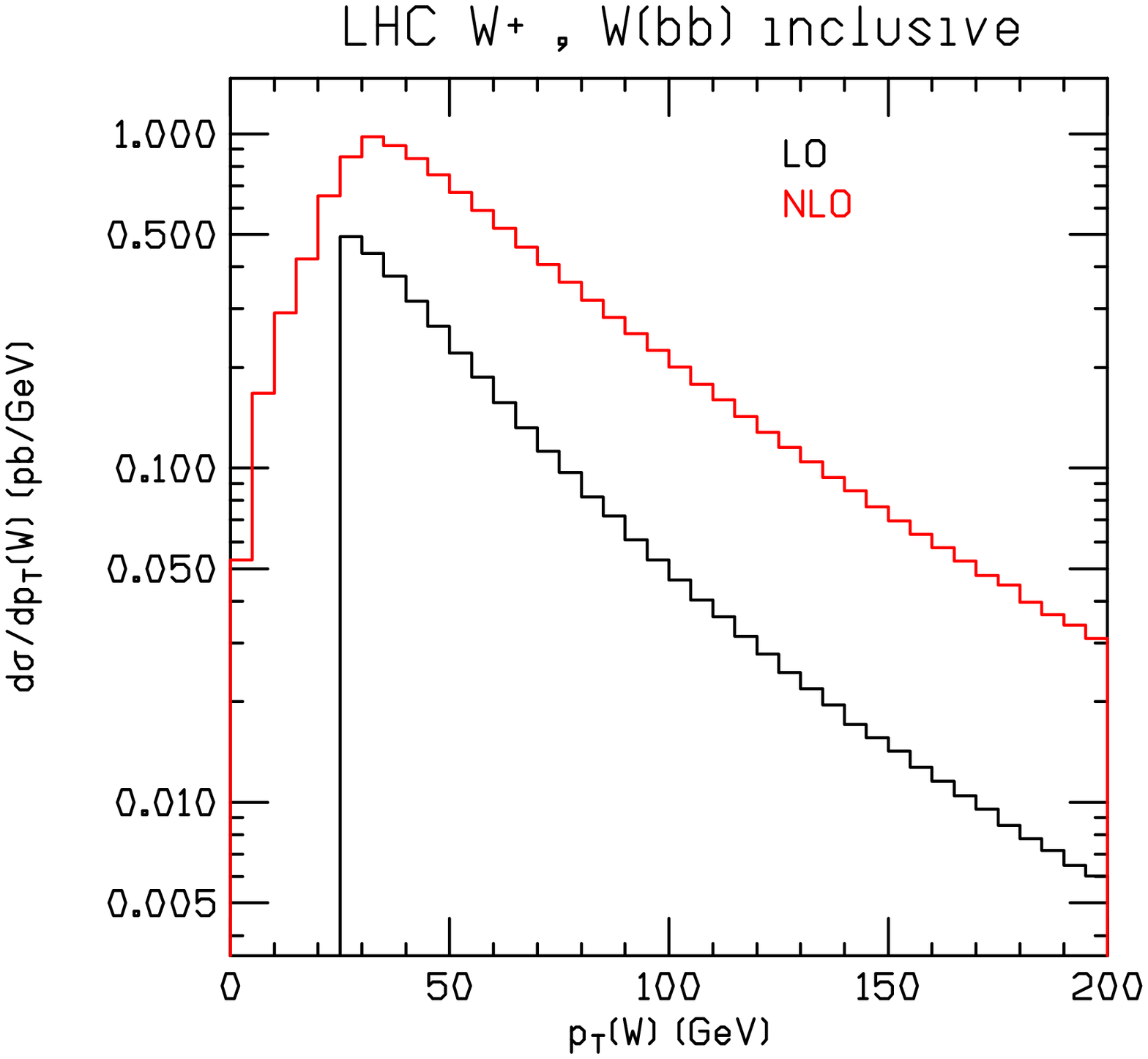}
\end{tabular}
\caption{$p_T(b)$ (upper plots) and $p_T(W)$ (lower plots)
  distributions for $W^+(bb)$ \emph{inclusive} production, at the Tevatron (l.h.s.\ plots)
  and at the LHC (r.h.s.\ plots).}
\label{fig:dist_signature2}
\end{center}
\end{figure}

In Figs.~\ref{fig:dist_signature3}--\ref{fig:dist_signature2} we show
the differential cross sections with respect to the transverse
momentum of the $b$ jet and of the $W$ boson, for $W^+b$
inclusive/exclusive and $W^+(b\bar{b})$ inclusive/exclusive
production. If there is more than one $b$ jet in the final state, the
$p_T$ of the highest-$p_T$ $b$ jet is chosen. We do not show
distributions for $W^-b$ and $W^-(b\bar{b})$ production at the LHC,
since they resemble the ones for $W^+b$ and $W^+(b\bar{b})$ production
at the LHC illustrated here.  For all final states, the NLO QCD
corrections modify the shape of both the $b$-jet and $W$-boson
transverse momentum distributions.

\section{Conclusions}\label{sec:conclusions}

In this paper we report on a NLO calculation of the production of a
$W$ boson with one $b$ jet.  We present both inclusive and exclusive
cross sections, as well as cross sections where the jet contains one
or two $b$ quarks.  We show that it is essential to keep the
$b$-quark mass finite throughout the calculation, and we are able to
overcome this technical hurdle.  The calculation is performed by
combining two previous NLO calculations of $Wb\bar b$
\cite{FebresCordero:2006sj} and $Wbj$ \cite{Campbell:2006cu}, taking
care to treat their overlap consistently.  The calculation that we
present thus represents the state of the art prediction
for $W$ plus one $b$ jet production at NLO in QCD.

These calculations can be compared with the large amount of data on
$W$ plus $b$ jets already gathered at the Tevatron, and soon to be
produced at the LHC.  They can also be compared with the inclusive
samples obtained by merging matrix-element calculations with parton
showers.

\section*{Acknowledgments}

\indent\indent We are grateful for conversations and correspondence
with Ann Heinson and Tony Liss. F.~M.\ and L.~R.\ thank the Aspen
Center for Physics for hospitality while this work was being
completed.  This work was supported in part by the U.~S.~Department of
Energy under contracts Nos.~DE-AC02-76CH03000, DE-FG02-91ER40662,
DE-FG02-91ER40677, and DE-FG02-97IR4102.  The work of D.~W.\ was
supported in part by the National Science Foundation under grants
NSF-PHY-0456681 and NSF-PHY-0547564.

\newpage


\end{document}